\documentclass[twocolumn, superscriptaddress, aps, pra, floatfix]{revtex4-2}

\usepackage{color}
\usepackage{graphicx}
\usepackage{dcolumn}
\usepackage{braket}
\usepackage{physics}
\usepackage{amsmath,amssymb}

\usepackage[breaklinks=true,colorlinks,citecolor=blue,linkcolor=blue,urlcolor=blue]{hyperref}
\usepackage{makeidx}

\usepackage[dvipsnames]{xcolor}
\usepackage{braket}

\usepackage[markup=underlined, defaultcolor=RedOrange]{changes}

\begin{document}

\author{Daniele Lamberto}
\affiliation{Dipartimento di Scienze Matematiche e Informatiche, Scienze Fisiche e  Scienze della Terra, Universit\`{a} di Messina, I-98166 Messina, Italy}

\author{Omar Di Stefano}
\affiliation{Dipartimento di Scienze Matematiche e Informatiche, Scienze Fisiche e  Scienze della Terra, Universit\`{a} di Messina, I-98166 Messina, Italy}

\author{Stephen Hughes}
\affiliation{Department of Physics, Engineering Physics and Astronomy, Queen’s University, Kingston, ON K7L 3N6, Canada}

\author{Franco Nori}
\affiliation{Theoretical Quantum Physics Laboratory, RIKEN Cluster for Pioneering Research, Wako-shi, Saitama 351-0198, Japan}
\affiliation{Physics Department, The University of Michigan, Ann Arbor, Michigan 48109-1040, USA}

\author{Salvatore Savasta}
\affiliation{Dipartimento di Scienze Matematiche e Informatiche, Scienze Fisiche e  Scienze della Terra,	Universit\`{a} di Messina, I-98166 Messina, Italy}

\newcommand{\figref}[1]{\mbox{Fig.~\ref{#1}}}
\newcommand{\figpanel}[2]{Fig.~\hyperref[#1]{\ref*{#1}(#2)}}
\newcommand{\figurepanel}[2]{Figure~\hyperref[#1]{\ref*{#1}(#2)}}
\newcommand{\figpanels}[3]{Figs.~\hyperref[#1]{\ref*{#1}(#2)-(#3)}}

\newcommand{\be}{\begin{equation}}
\newcommand{\ee}{\end{equation}}
\newcommand{\bea}{\begin{eqnarray}}
\newcommand{\eea}{\end{eqnarray}}

\renewcommand{\eqref}[1]{\mbox{Eq.~(\ref{#1})}}
\newcommand{\eqaref}[1]{\mbox{Equation~(\ref{#1})}}

\title{Quantum Phase Transitions in Many-Dipole Light-Matter Systems}

\begin{abstract}

A potential phase transition between a normal ground state and a photon-condensed ground state in many-dipole light-matter systems is a topic of considerable controversy, exasperated by  conflicting no-go and counter no-go theorems and often ill-defined models. We clarify this long-lasting debate by analyzing two specific arrangements of atoms, including a 3D cubic lattice and a cavity-embedded square lattice layer---which provides a physical model for single-mode cavity QED with coupled dipoles in the thermodynamic limit. These models are shown to significantly differ from the standard Dicke model and, in the thermodynamic limit, give rise to  renormalized Hopfield models. We show that a ferroelectric phase transition can (in principle) still occur and the description of the abnormal phase beyond the critical point requires the inclusion of nonlinear terms in the Holstein-Primakoff mapping. 
We also show how our model
agrees with recent experiments.

\end{abstract}

\maketitle

\begin{figure}[b]
    \centering
    \begin{minipage}{0.48\linewidth}
        \centering
        \includegraphics[width=\linewidth]{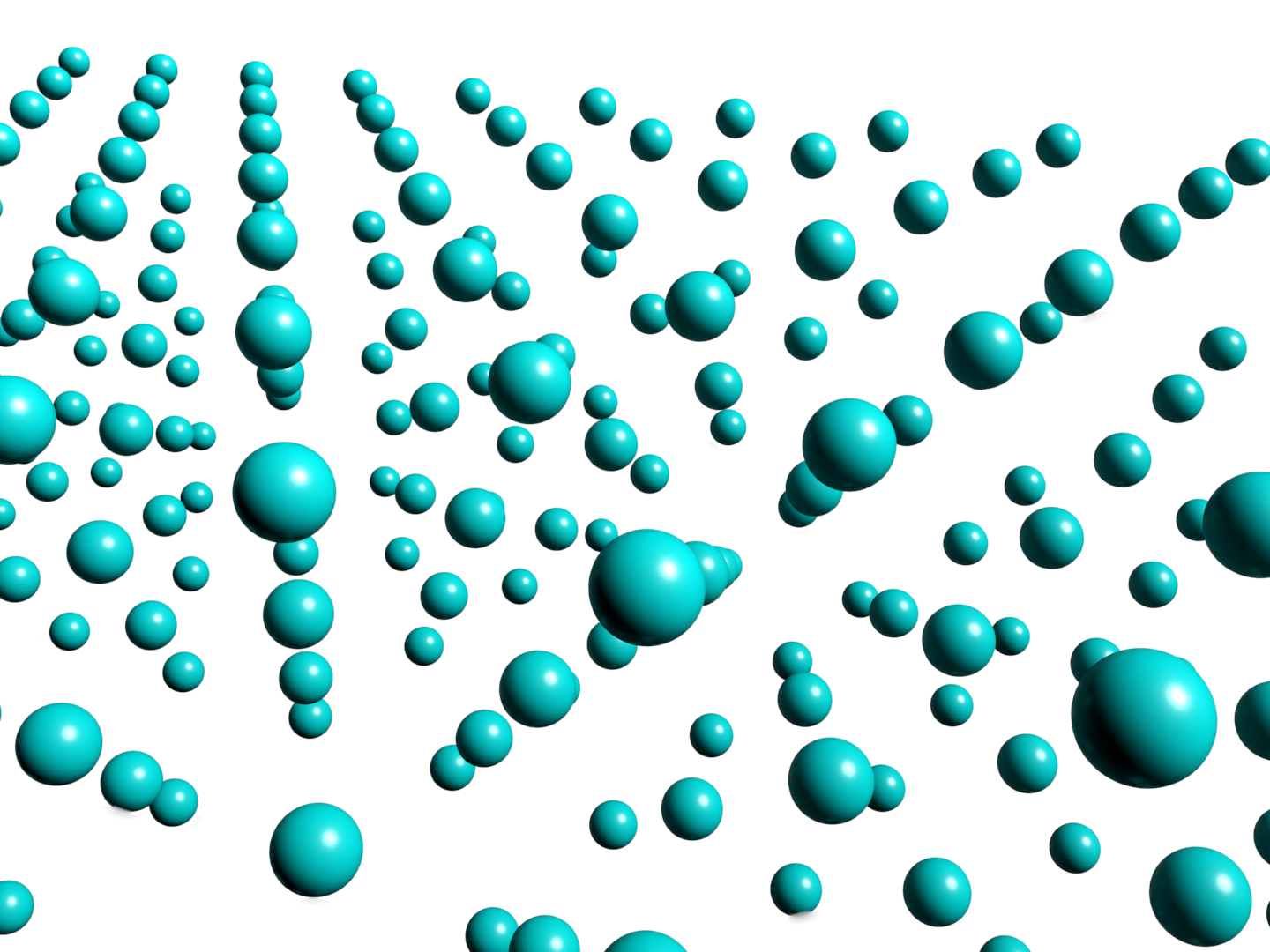}
        \label{fig:bulk}
    \end{minipage}\hfill
    \begin{minipage}{0.48\linewidth}
        \centering
        \includegraphics[width=\linewidth]{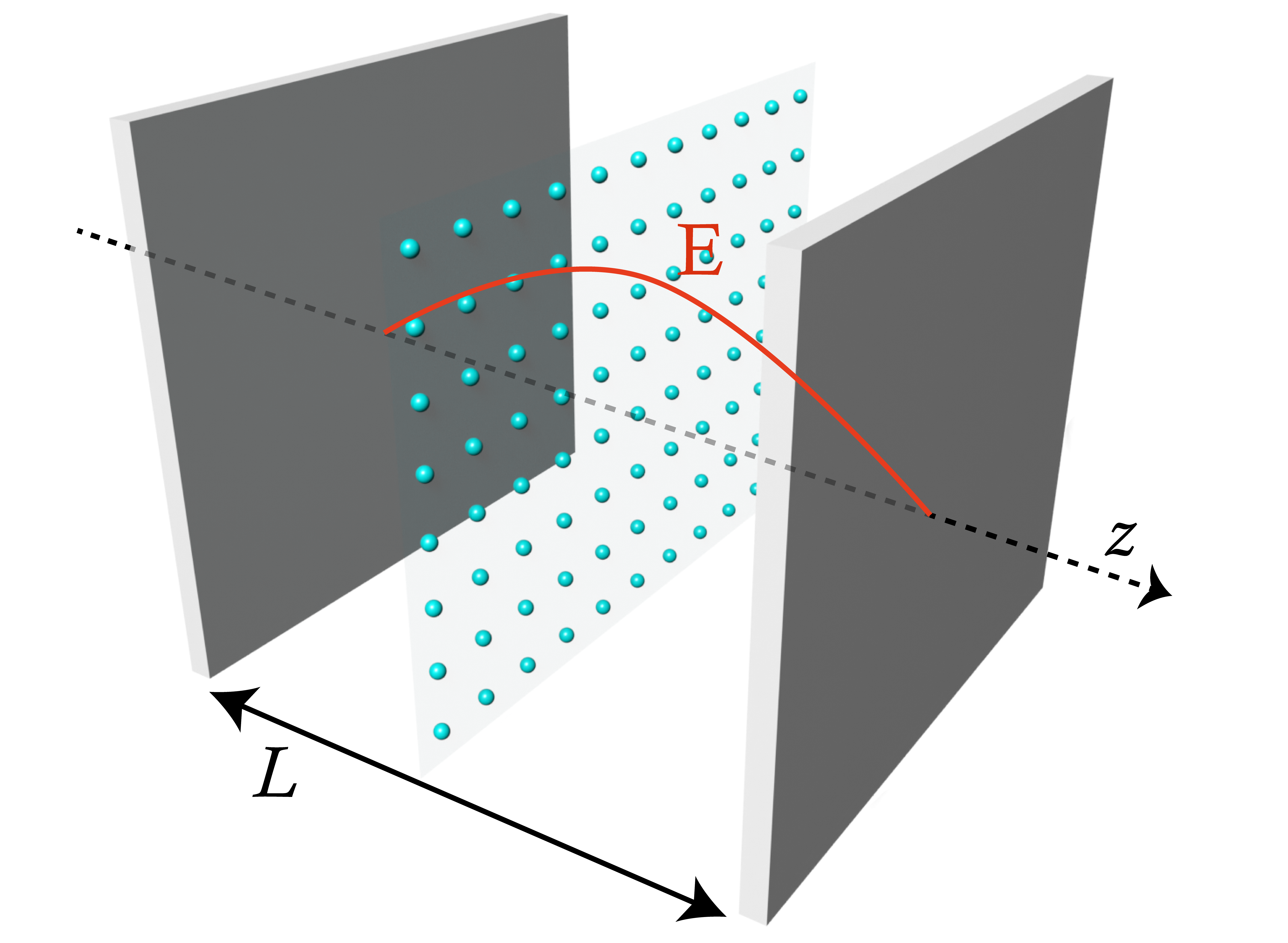}
        \label{fig:layer}
    \end{minipage}
    \caption{Schematic of atoms arranged in a 3D lattice (left) and in a cavity-embedded planar lattice (right). In the latter case, we consider radiation modes orthogonal to the atomic layer ($xy$ plane).}
    \label{fig:systems}
\end{figure}


The possibility of a phase transition between a normal state and a photon condensate state within light-matter systems under the influence of electric dipolar interactions, commonly referred to as superradiant phase transition (SPT), has been a long-standing debate for many decades \cite{Dicke54, Hepp_Lieb1973, Mallory1969, Wang1973, Hioe73, Zakowicz1975, Yamanoi1976, Bamba2014, Rabl2016, Debernardis2018, Debernardis_Deliberato2018, Ashida20_PRX}.  The superradiant phase is characterized by a macroscopically large number of coherent photons in the ground state. Even in recent years, several seemingly contradictory no-go and counter no-go theorems still dispute over the occurrence of a SPT in such systems \cite{Keeling2007, Knight1978, Mazza2019, Andolina2019, Lenk2020, Savasta2022}. Notably, such debates are limited to the case of electric dipolar interactions, as there is a general consensus regarding the potential occurrence of a SPT in the presence of magnetic interactions, given that in these systems the so-called ${\bf A}^2$ term in the Coulomb gauge Hamiltonian (or ${\bf P}^2$ term in the multi-polar gauge Hamiltonian), which prevents the SPT, can be significantly smaller or possibly absent \cite{Polini2020, Zueco2021}. 
However, these quadratic terms are not only required to ensure gauge-invariance, but they are necessary to 
recover fundamental classical limits \cite{Garziano2020,schafer2020,Hughes2024-Optica}. 

Very recently,  a Dicke-like SPT has been observed in a magnetic system where a magnon mode of ordered Fe$^{3+}$ spins interacts with paramagnetic Er$^{3+}$ spins \cite{kim2024observation}.
Recent no-go theorems have presented rather general demonstrations that gauge invariance forbids any phase transition to a photon condensate state when the cavity-photon mode is assumed to be spatially uniform in the region where the dipoles are located \cite{Knight1978,Andolina2019}, or when magnetic interactions can be neglected \cite{Savasta2022}. However, conversely, a number of papers still support the plausibility of a SPT in many-dipole cavity-QED systems, grounded on the assumption that the system's Hamiltonian can be mapped onto a Dicke-like model \cite{Keeling2007, Domokos2014, Stokes2020}. 

This apparent possibility originates from the interplay of direct electrostatic interactions and a transverse matter field term, resulting in reciprocal compensation within the multipolar gauge.
There 
exists documented instances where 
this compensation has been effectively utilized, and 
outcomes have been derived \cite{Andrews, Bradshaw}. 
In contrast, Ref.~\cite{Bamba2014} predicts that 
longitudinal dipole-dipole interactions do not enable any quantum phase transition (QPT), at least for the infinite, homogeneous, and isotropic system of non-overlapping dipoles.
Moreover, there is no general agreement on the nature of this controversial QPT. According to \cite{Keeling2007}, the phase transition occurring in Dicke-like models corresponds  to a spontaneous polarization of the two-level systems, which does not however lead to a spontaneous transverse electric field, i.e., the QPT is ferroelectric-like. 
However, Ref.~\cite{Vukics2012} argues that a mean field occupation in a mode of the transverse displacement field, far from the dipoles, implies a mean field in the transverse electric field; Refs.~\cite{Stokes2020, Nazir_revmodphy} 
imply the nature of the QPT is gauge dependent and 
purely ferroelectric only in the Coulomb gauge.

To help solve these long-standing controversies, we investigate the interaction between the electromagnetic field and an ordered lattices of atoms. Specifically, we consider  isotropic, localized two-level atomic dipoles with threefold orientation degeneracy.  
When applying the two-level approximation, a consistent model of the atom requires one to consider the orientation degeneracy in one of the two levels (e.g., the excited state), since the electric-dipole transition is allowed only between states with different parities (e.g., between $s$ and $p$ orbitals). 
Considering these 
lattices, we take into account (in a simple and rigorous way) the discrete and non-overlapping system topology, which is often overlooked during the different  approximations and limits employed to derive the Dicke model.
We start by considering a cubic lattice array (an isotropic system in the long wavelength limit) and successively a cavity-embedded planar layer with a square lattice (see Fig.~\ref{fig:systems}), which provides a physically motivated model for single-mode cavity QED in the dipole approximation and in the thermodynamic limit.



{\em 3D Lattice.---}We consider a three-dimensional lattice of atoms interacting with an electromagnetic field. 
The radiation field is quantized by the introduction of radiation bosonic operators $a_{{\bf k}, \lambda}$ for each mode $\bf k$ and polarization $\hat{\bf e}_\lambda$. We denote the field frequency with $\omega_k = v \left|{\bf k} \right|$, where $v=c / \sqrt{\epsilon_m}$ is the speed of light and $\epsilon_m$ is the dielectric constant of the surrounding medium. 
In contrast, the atoms are modeled as localized charges in the sites ${\bf R}_n$ of a lattice. In the long-wavelength approximation, the total polarization density is thus expressed as ${\bf P(r)} = \sum_n {\bf d}_n \delta ({\bf r} - {\bf R}_n)$, where ${\bf d}_n$ is the total electric dipole of the \textit{n}-th atom, thus assuming non-overlapping dipoles for different atoms. The usual electrostatic dipole-dipole interaction is given by
\begin{equation}
    H_{\rm dip} = \frac{1}{8 \pi \epsilon_0 \epsilon_m} \sum_{n \neq m} \frac{{\bf d}_n \cdot {\bf d}_m - 3 ({\bf d}_n \cdot \hat{\bf r}_{nm})({\bf d}_m \cdot \hat{\bf r}_{nm})}{r_{nm}^3} \, , \label{eq:H_dip}
\end{equation}
where ${\bf r}_{nm} = {\bf R}_n - {\bf R}_m$ and $\hat{\bf r}_{nm} = {\bf r}_{nm} / r_{nm}$ is the associated unit vector. We then perform the two-level approximation (to derive Dicke-like models), reducing the atomic states to just the ground and excited levels, where we consider the excited states to have a threefold orientation degeneracy (see App.\, \ref{sec:two_lev}). 
Next, we bosonify the system in the thermodynamic limit (finite density as $N, \, V \to \infty$) through the use of generalized Holstein-Primakoff (HP) transformations (see App.\,\ref{sec:generalized_HP}) \cite{Brandes03, Brandes03_PRL}. 
The ensuing Hamiltonian for the matter system (according to the usual definition in condensed matter physics) is constituted by the  array of dipoles with their electrostatic interactions:
\begin{eqnarray}
    H_{\rm mat} & = & \hbar \omega_0 \sum_{\alpha, {\bf k}} b^\dag_{{\bf k}, \alpha} b_{{\bf k}, \alpha} + \hbar \sum_{\alpha, \beta, {\bf k}} \chi^2 \omega_0 f_{{\bf k}, \alpha, \beta} \times \nonumber \\
    & & \left[ \left( b_{{\bf k}, \alpha} + b^\dag_{-{\bf k}, \alpha} \right) \left( b_{-{\bf k}, \beta} + b^\dag_{{\bf k}, \beta} \right) \right] \, , \label{eq:H_mat}
\end{eqnarray}
where the second term is the electrostatic dipole-dipole interaction in \eqref{eq:H_dip}, $f_{{\bf k}, \alpha, \beta}\approx \left( 3 (\hat{\bf k} \cdot \hat{{\bf e}}_\alpha) (\hat{\bf k} \cdot \hat{{\bf e}}_\beta) - \delta_{\alpha \beta}\right) / 3$ for a simple cubic lattice in the long-wavelength limit and  $\chi = \sqrt{{d^2 N} / {2 \hbar \epsilon_0 \epsilon_m V \omega_0}}$ (see App.\,\ref{sec:dipolar_int}) \cite{Cohen_Keffer1955}. 
We can diagonalize the matter Hamiltonian through a Bogoliubov transformation, which leads to $H_{\rm mat} = \hbar \sum_{\alpha, {\bf k}} \Tilde{\omega}_{{\bf k}, \alpha} c^\dag_{{\bf k}, \alpha} c_{{\bf k}, \alpha} $, with renormalized matter frequency $\Tilde{\omega}_{{\bf k}, \alpha} = \omega_0 \sqrt{1 + 4 \eta^2 f_{{\bf k}, \alpha, \alpha}}$ and  bosonic eigenoperators $ c_{{\bf k}, \alpha}$.
Here, we used the relation $\chi = \eta$, \textit{valid only in the 3D case}, where $\eta$ is the light-matter coupling, as discussed below.
The new transverse eigenmodes describe the collective matter excitations which effectively couple with the radiation field, giving rise to a Hopfield-like model. The light-matter Hamiltonian can be obtained by applying the minimal coupling replacement in the Coulomb gauge, or, equivalently, applying a suitable unitary transformation \cite{Garziano2020} to the photonic Hamiltonian $H_{\rm ph} = \hbar \sum_{\lambda, {\bf k}} \omega_k a^\dag_{{\bf k}, \lambda} a_{{\bf k}, \lambda}$ in the multipolar gauge. This procedure leads to a Hopfield-like Hamiltonian (see App.\,\ref{sec:derivation_H}), which in the multipolar gauge is given by
\begin{equation}
    H = H_{\rm ph} + H_{\rm mat} + H_{\rm I_1} + H_{\rm I_2} \, , \label{eq:H_M_bulk}
\end{equation}
with the light-matter interaction terms
\begin{eqnarray}
    H_{\rm I_1} \!\! &{=}& \!\! i \hbar \sum_{\alpha, \lambda, {\bf k}} \! \eta' \sqrt{\omega_k \Tilde{\omega}^\perp_{\bf k}} \! \left( a^\dag_{-{\bf k}, \lambda} {-} a_{{\bf k}, \lambda} \right) \!\! \left( c^\dag_{{\bf k}, \alpha} {+} c_{-{\bf k}, \alpha} \right) e_{\lambda_{\alpha}} \nonumber \\
    H_{\rm I_2} \!\! &{=}& \!\! \hbar \!\!\! \sum_{\alpha, \beta, \lambda, {\bf k}} \!\!\! \eta'^2 \Tilde{\omega}^\perp_{\bf k} \! \left( c^\dag_{{\bf k}, \alpha} {+} c_{-{\bf k}, \alpha} \right) \!\! \left( c^\dag_{-{\bf k}, \beta} {+} c_{{\bf k}, \beta} \right) \! e_{\lambda_{\alpha}} \! e_{\lambda_{\beta}}, 
\end{eqnarray}
where $\Tilde{\omega}^\perp_{\bf k} = \omega_0 \sqrt{1 + 4 \eta^2 f^\perp_{\bf k}}$ and $\eta' = \eta\, \omega_0/ \Tilde{\omega}^\perp_{\bf k}$ are the transverse resonance frequency and the coupling renormalized by the dipole-dipole interaction.
Furthermore, $f^\perp_{\bf k}$ and $f^\|_{\bf k}$ represent the transverse and the longitudinal part of $f_{{\bf k}, \alpha, \alpha}$, respectively, while we defined $e_{\lambda_{\alpha}} = \hat{{\bf e}}_\lambda \cdot \hat{{\bf e}}_\alpha$, where the $\{ \hat{{\bf e}}_\alpha \}$ are a generic set of orthonormal vectors chosen as a basis for the electric dipole orientation. 
The dispersion relations for the transverse modes can be obtained by diagonalizing the Hopfield-like Hamiltonian (\ref{eq:H_M_bulk}), resulting in (see App. \ref{sec:disp_rel})
\begin{equation}
    \frac{\omega^2_{\bf k}}{\Omega^2_\perp} = 1 + \frac{ 4 \eta'^2 \Tilde{\omega}_{\bf k}^{\perp^2}}{ \Tilde{\omega}_{\bf k}^{\perp^2} - \Omega^2_\perp }\, , \label{eq:disp_rel_bulk}
\end{equation}
where $\hbar \Omega_\perp$ is the energy of the transverse polaritons. The longitudinal modes are not affected by the interaction with photons and their energy is already in diagonal form in 
\eqref{eq:H_M_bulk}:  $\Omega_\| = \Tilde{\omega}^\|_{\bf k}$. We also observe that the same dispersion relations are obtained by diagonalizing the full light-matter Hamiltonian, without performing the initial Bogoliubov transformation on the matter subsystem (see App.\,\ref{sec:disp_rel}). Moreover, the same results can be obtained in the Coulomb gauge, if gauge invariance is treated with a consistent approach \cite{DiStefano2019, Garziano2020, Savasta-Nori2021,Hughes2023}. Notice that dispersion relation in \eqref{eq:disp_rel_bulk}, obtained by diagonalizing the Hopfield model, also agree with the results obtained solving the corresponding Maxwell equations.

Let us now discuss the possibility for the system to undergo a QPT. 
\begin{figure}[b]
    \centering
    \includegraphics[width=0.49\textwidth]{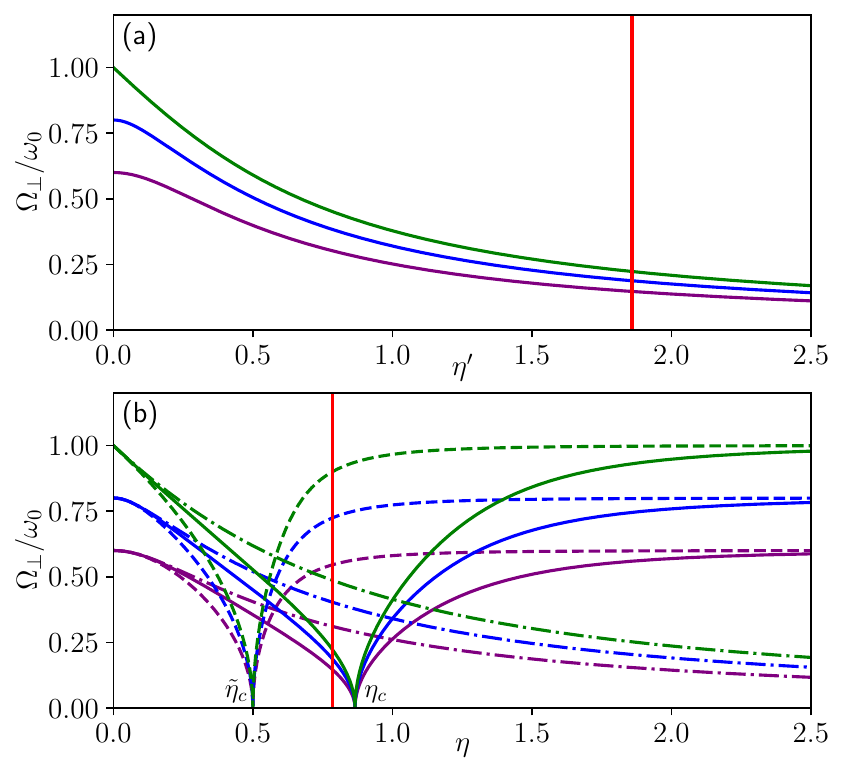}
    \caption{Lower polariton resonances for different modes $\omega_k / \omega_0 = 0.6$ (purple), $0.8$ (blue), $1$ (green). The vertical red line indicates the coupling strength $\eta' = 1.83$ measured in \cite{Nature_exp_nanoparticles}, corresponding to $\eta = 0.78$. (a): Lower polaritons as functions of $\eta'$. (b): Comparison of lower polaritons obtained from renormalized Hopfield-like model (solid lines), Dicke-like model (dashed lines) and Hopfield model without dipole-dipole interactions (dash-dotted lines) versus $\eta$.}
    \label{fig:disp_curves_vs_beta}
\end{figure}
\eqaref{eq:disp_rel_bulk} is derived by a renormalized Hopfield-like model with effective coupling constant $\eta'$, thus, as it is well- known, it does not admit any radiation induced QPT (see Fig.~\ref{fig:disp_curves_vs_beta}a) \cite{NatafCiuti2010}.
On the other hand, the Hamiltonian for the matter subsystem in \eqref{eq:H_mat}  enables the occurrence of a QPT, as testified by the softening of the transverse renormalized matter frequency $\Tilde{\omega}^\perp_{\bf k}$ at increasing dipole-dipole interaction strengths $\eta$ (notice that $f^\perp_{\bf k} <0$). This ferroelectric QPT occurs even without including  {retardation effects} and gives rise to a transverse polarization condensate $\langle P^\perp_{\bf k} \rangle \neq 0$ (see App.\,\ref{sec:aboveSPT}).
In particular, for isotropic systems in the long-wavelength approximation, the QPT can be achieved for  $\eta> \eta_c \approx \sqrt{3}/2$. In contrast, near a QPT, $\eta' = \eta\, \omega_0/ \Tilde{\omega}^\perp_{\bf k}$ is not a suitable parameter given that it diverges for $\eta \to \eta_c$. This relation between $\eta$ and $\eta'$  explains why the Hopfield model remains a valid description even for many-dipole systems approaching a ferroelectric QPT.

It is worth noticing that neither the ground-state transverse polarization, nor the critical point are affected by the interaction with the photon field, as expected by a ferroelectric phase transition.
However, this QPT, involving transverse matter excitations, reasonably also affects  transverse polaritons.
We can calculate the transverse dispersion in the condensed phase using higher-order terms in the HP mapping, obtaining (see App.\,\ref{sec:aboveSPT})
\begin{equation}
    \frac{\omega^2_{\bf k}}{\Omega^2_\perp} = 1 + \frac{ \omega^2_0 / f^\perp_{\bf k} }{ \omega^2_0 \left( 1 - 16 \eta^4 {f_{\bf k}^\perp}^2 \right) + \Omega^2_\perp }\,, \quad \text{for}\: \, \eta > \eta_c\,, \label{eq:disp_rel_bulk_transv}
\end{equation}
while Eq.~(\ref{eq:disp_rel_bulk}) is valid for $\eta < \eta_c$. In particular, the ferroelectric condensation of the dipoles system determines a macroscopic transverse polarization density,
which is not affected by the interaction with the photon field, but
induces a macroscopic occupation of $\langle D_{\bf k} \rangle \neq 0$, which, in the multipolar gauge, is proportional to the field momentum, implying $\langle a_{{\bf k},\lambda} \rangle \neq 0$. This occurrence 
caused 
recent works \cite{Stokes2020, Nazir_revmodphy} to characterize the phase transition as ferroelectric or superradiant depending on the gauge choice. 
In contrast,
we obtain $\langle D_{\bf k} \rangle = \langle P^\perp_{\bf k} \rangle$, which, from the definition of the displacement field, implies $\langle E^\perp_{\bf k} \rangle = 0$. 
We observe that quantities as $D_{\bf k}$ and the transverse electric field $E^\perp_{\bf k}$ are physical quantities, whose expectation values are not gauge dependent. However, operators
such as the field momentum and the photon creation and destruction operators are nonphysical quantities which can be used only as calculation tools \cite{Scully_book}, and are not suitable to characterize  physical processes.


We now briefly discuss a potential cancellation procedure which is used to justify the usual Dicke model \cite{Yamanoi_book, Keeling2007, Stokes2020, Nazir_revmodphy}. 
It is based on the interplay between electrostatic terms in \eqref{eq:H_mat} (the dipole-dipole interaction) and the inter-atomic part of $H_{\rm I_2}$ (which originates from the light-matter interaction). 
After compensating these two terms, to obtain the Dicke model, a single-polarization single-mode approximation for the radiation and a strict two-level approximation for the dipoles are employed. The intra-atomic part of $H_{\rm I_2}$, in the strict two-level approximation, becomes proportional to the identity operator. However, this term results from the combination of  electrostatic and transverse contributions and it may not be appropriate for electrostatic interactions to neglect the orientation degree of freedom. 
Following this procedure, the Dicke-like multipolar-gauge Hamiltonian for a single mode is obtained (e.g. Ref.~\cite{Stokes2020}): 
\begin{equation}
    \tilde H = \hbar \omega_0 b^\dag b + \hbar \omega_k a^\dag a - i \hbar \eta \sqrt{\omega_k \omega_0} \left( a^\dag - a \right) \left( b^\dag + b \right) \,. \label{eq:H_M_canc_bulk}
\end{equation}
This model, in contrast with the previous one, predicts a light-induced QPT (superradiant QPT, SPT), for a critical value of the coupling $\Tilde{\eta}_c = 0.5$ \cite{Brandes03, Stokes2020}.

\begin{figure}[t]
    \centering
    \includegraphics[width=0.49\textwidth]{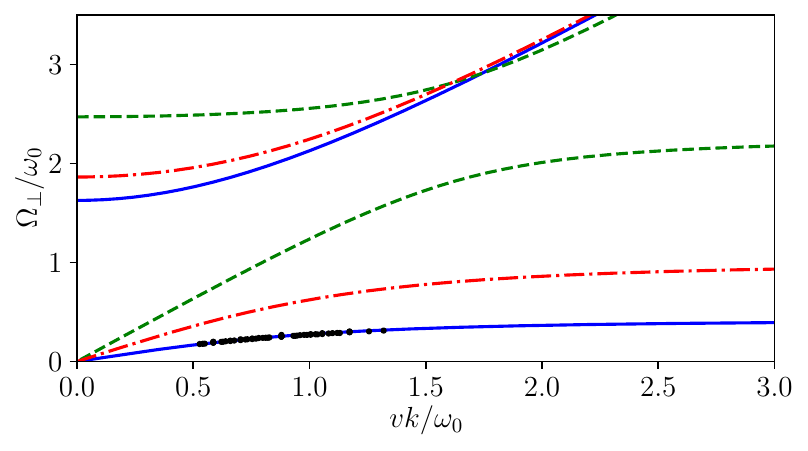}
    \caption{Comparison of transverse dispersion curves obtained for the  renormalized Hopfield-like model [\eqref{eq:disp_rel_bulk}] (blue solid), Dicke-like model (green dashed) and Hopfield model without dipolar interactions (red dash-dotted). Coupling constant $\eta' = 1.83$. Experimental data (black dots) from Ref. \cite{Nature_exp_nanoparticles} are included, showing a strong agreement with the model proposed.}
    \label{fig:disp_curves}
\end{figure}

Figure \ref{fig:disp_curves_vs_beta}b shows a softening of the lower polariton modes obtained from the Hamiltonians in Eqs.~(\ref{eq:H_M_bulk}) and (\ref{eq:H_M_canc_bulk}), until they all go to zero for the critical values of the coupling $\eta_c$ and $\Tilde{\eta}_c$, respectively, which is a signature of the presence of a quantum phase transition. In contrast, the polariton modes calculated from a pure Hopfield model in which the dipolar interaction is not taken into account do not present any kind of QPT. 
In \figref{fig:disp_curves}, we compare the dispersion curves for the transverse and longitudinal sectors in both Hopfield-like [\eqref{eq:H_M_bulk}] and Dicke-like [\eqref{eq:H_M_canc_bulk}] models. In particular, Fig.~\ref{fig:disp_curves} shows the theoretical predictions of both models and the experimental data taken from Ref.~\cite{Nature_exp_nanoparticles}. The data set reports measurements of the lower polariton branch of a three-dimensional artificial crystal made of spatially separated gold nanoparticles (see App.\,\ref{sec:experimental_data} for details) \cite{Lamowski2018}.
The nanoparticles, each supporting triply-degenerate localized dipolar surface plasmons, couple through dipole-dipole interactions, giving rise to collective plasmons that extend over the whole metamaterial. These excitations can be described in terms of collective bosonic operators, analogously to the atomic collective excitations (in the thermodynamic limit) that we considered above. 

These artificial gold crystals can reach very high light-matter interaction strengths. Hence, they represent an ideal testbed for discerning among the different models considered, and help solve the long-standing debate about QPTs in many-dipoles systems. The measurements in Fig. \ref{fig:disp_curves} have been already compared in Ref.~\cite{Nature_exp_nanoparticles} with a renormalized Hopfield model, showing excellent agreement.
The high coupling achieved in the experiment ($\eta' = 1.83$, corresponding to $\eta= 0.78 > \tilde \eta_c$) would imply a condensed phase according to the Dicke-like model. The Dicke dispersion relation, shown in Fig.~\ref{fig:disp_curves}, strongly differs from the data. However, we point out that, beyond the critical point, the dispersion relation for the atomic system (shown in Fig.~\ref{fig:disp_curves}) and for the artificial crystal  may present quantitative differences owing to the different nonlinear response of these systems. It is worth noticing that the achieved coupling strength $\eta = 0.78$ is not far from $\eta_c \sim 0.87$ (see the red vertical line in Fig.~\ref{fig:disp_curves_vs_beta}), and the consequent softening of $\tilde \omega_{{\bf k} \perp}$ can be appreciated. We observe that also the Hopfield model without dipole-dipole interactions fails to reproduce correctly the data. 

{\em 2D lattice.---}Next, we consider a system composed of a two-dimensional layer of atoms, identified as the $xy$ plane, interacting with a radiation field confined in an ideal cavity (e.g., with dielectric mirrors) (Fig.~\ref{fig:systems}b).
This planar configuration naturally induces a decomposition of the vector potential in terms of ${\bf k}_\|$, the in-plane discrete component of the wave vector with corresponding quantization surface $S$, and its orthogonal component $k_z$, quantized by the length of the cavity $L$. 
Considering radiation modes with wavevectors orthogonal to the planar surface (${\bf k} = k_z {\bf \hat{z}}$), the treatment considerably simplifies given that the two light polarization vectors $\hat{{\bf e}}_\lambda$ now lie in the $xy$ plane and the radiation bosonic operators become independent on ${\bf k}_\|$, i.e., $a_{k_z, \lambda}$. The Hamiltonian of the free electromagnetic field for such system is \cite{Savasta1996}
\begin{equation}
    H_{\rm ph} = \sum_\lambda \sum_{j=e,o} \sum_{k_z > 0} \hbar \omega_k  a^\dag_{j, k_z, \lambda} a_{j, k_z, \lambda} \, , \label{eq:H_ph_layer}
\end{equation}
where we defined the even and odd radiation modes operators $a_{e(o), k_z, \lambda} =  \left( a_{k_z, \lambda} \pm a_{- k_z, \lambda} \right)/\sqrt{2}$.
Following the procedure for the 3D dipoles lattice, we firstly perform the two-level approximation taking into account the dipole orientations, and successively construct two-dimensional collective bosonic operators $b_{\bf k_\|}$. For incidence orthogonal to the planar surface (${\bf k}_\| = 0$), the matter Hamiltonian reads
\begin{equation}
    H_{\rm mat} {=} \hbar \omega_0 \sum_{\alpha} b^\dag_{\alpha} b^{\phantom{\dag}}_{\alpha} {+} \hbar \sum_{\alpha, \beta} \chi^2 \omega_0 f_{{\bf z}, \alpha, \beta} \! \left( b^{\phantom{^\dag}}_{\alpha} + b^\dag_{\alpha} \right) \!\! \left( b^{\phantom{\dag}}_{\beta} + b^\dag_{\beta} \right) \label{eq:H_mat_layer}
\end{equation}
where $b_{\alpha} \equiv b_{{\bf k_\|}=0, \alpha}$ and the structure-dependent factor for the 2D square lattice is $\chi = \sqrt{d^2 \mu / \hbar \epsilon_0 \epsilon_m a^3 \omega_0}$, with $a$ the lattice constant and $\mu \approx 6.78 / 4 \pi$ (see App.\,\ref{sec:dipolar_int}).

As in the 3D lattice case, the Hamiltonian in \eqref{eq:H_mat_layer} can be diagonalized through the introduction of eigen-operators $c_\alpha$, with a corresponding renormalized frequency $\Tilde{\omega}_{\alpha} = \omega_0 \sqrt{1 + 4 \chi^2 f_{{\bf z}, \alpha, \alpha}}$.
Therefore, the light-matter Hamiltonian in multipolar gauge reads
\begin{equation}
    H = H_{\rm ph} + H_{\rm mat} + H_{\rm I_1} + H_{\rm I_2} \, , \label{eq:H_M_layer}
\end{equation}
with the light-matter interaction terms
\begin{eqnarray}
    \!\!\! H_{\rm I_1} \!\! & = & \!\! i \hbar \!\!\! \sum_{\lambda, k_z>0} \!\!\!\! \eta' \sqrt{ \omega_{k_z} \Tilde{\omega}^\perp} \left( a^\dag_{e, k_z, \lambda} - a^{\phantom{\dag}}_{e, k_z, \lambda} \right) \left( c^\dag_{\lambda} + c^{\phantom{\dag}}_{\lambda} \right) \nonumber \\
    \!\!\! H_{\rm I_2} \!\! & = & \!\! \hbar \!\!\! \sum_{\lambda, k_z>0} \!\!\!\! \eta'^2 \Tilde{\omega}^\perp \left( c^\dag_{\lambda} + c^{\phantom{\dag}}_{\lambda} \right) \left( c^\dag_{\lambda} + c^{\phantom{\dag}}_{\lambda} \right) \, ,
\end{eqnarray}
where $\eta' = \eta \, \omega_0 / \Tilde{\omega}^\perp$ is the renormalized light-matter coupling for the planar layer, with $\eta = \sqrt{d^2 N / \hbar \epsilon_0 \epsilon_m S L \omega_0}$ and $\Tilde{\omega}^\perp$ renormalized transverse matter frequency. We have chosen the same basis for the dipole orientation and the radiation polarization. When considering a single $k_z$ and a single transverse polarization mode, the Hamiltonian (\ref{eq:H_M_layer}) reduces to that of a two coupled harmonic oscillators model, as the Dicke Hamiltonian in the thermodynamic limit, but with the presence of the so-called ${\bf P}^2$ term ($H_{\rm I_2}$). 

From the total Hamiltonian in (\ref{eq:H_M_layer}), we can derive the transverse dispersion relation:
\begin{equation}
    \frac{\Omega^2_\perp - \Tilde{\omega}^{\perp^2}}{2 \Tilde{\omega}^\perp} + 2 \eta'^2 \Tilde{\omega}^\perp \sum_{k_z > 0} \frac{\Omega^2_\perp}{\omega^2_{k_z} - \Omega^2_\perp} = 0 \, . \label{eq:disp_rel_transv_layer}
\end{equation}
In the single-mode approximation, such a relation reduces to the dispersion relation derived from a renormalized Hopfield-like model, as in the 3D lattice.
We notice that in both cases the dipole-dipole structure factor $\chi$ depends on $a^{-3}$. However, the 2D and 3D lattices exhibit a different scaling of the light-matter coupling with respect to the lattice constant (see App.~\ref{sec:dipolar_int}). While in the 3D lattice we can relate $\eta$ to the volumetric density $\rho$, 
(leading to the relation $\eta^2 \propto \rho \propto a^{-3}$),
in the 2D lattice this proportionality cannot be established anymore since the charges are now ordered in a planar structure with a surface density $\sigma \propto a^{-2}$ (which appears in the light-matter coupling $\eta$ together with the cavity length $L$). Thus, we cannot simply relate the light-matter coupling $\eta$ to $\chi$, which we remark is the factor responsible for the QPT.
Moreover, we point out that in this case the cancellation procedure loses meaning given that the couplings governing the dipole-dipole and the light-matter interactions, $\chi$ and $\eta$ respectively, are different, although both depend on the atomic dipole moment.
These evidences further confirm the ferroelectric nature of the QPT.

{\em Discussion and conclusions.---}We have shown that the standard Dicke model, a widespread  description of many-dipole cavity-QED systems (in the dipole and single-mode approximations), is not a suitable description for simple systems of  two-level atomic dipoles with the usual orientation degeneracy.
The proper description for systems of non-overlapping dipolar quantum emitters is provided by the Hopfield model, renormalized to include electrostatic dipole-dipole interactions.
Our analysis agrees with recent experimental results on artificial crystals made of gold nanoparticles, which display analogous linear optical properties (in agreement with classical Maxwell solutions).
Although the Hopfield model does not admit any radiation-induced QPT, the system of dipolar quantum emitters can still undergo, at least in principle, a QPT when the dipole-dipole interaction strength reaches the critical value $\chi_c = \sqrt{3}/2$ (for a 3D lattice of atoms), different from the value predicted by the corresponding Dicke-like model.

The potential QPT determines the occurrence of a macroscopic transverse matter polarization field in the system ground state which affects the interaction of the matter system with light, hence modifying the dispersion relations for transverse polaritons in the normal and ferroelectric phases. 
Our results clarify the ferroelectric nature of the predicted QPT.
Although possible in principle, the ferroelectric QPT is likely very difficult to realize. Owing to the small size of the fine structure constant, $\chi \approx 10^{-3} \, r^2 \lambda_0 / a^3$ is usually far from the critical value $\chi_c$, with $\lambda_0 = 2 \pi v / \omega_0$ being the matter wavelength and $r$ the mean atomic radius. Notice, also, that $4 r < a$ is typically required to reasonably avoid overlap of the emitters wavefunctions.
It would be interesting to extend this analysis to different lattices of natural and artificial atoms, especially to systems of anisotropic quantum emitters with removed orientation degeneracy.

\bibliographystyle{ieeetr}
\bibliography{biblio}


\onecolumngrid
\appendix
\newpage

\section{Two-level approximation with threefold degeneracy} \label{sec:two_lev}

In the derivation of the Hamiltonians presented in this paper, we reduce the atomic states to two levels. Here we consider isotropic atoms with inversion symmetry. Owing to isotropy and electric-dipole selection rules,
a consistent model requires that one of the two levels (we chose the excited state) is threefold degenerated, while the ground state is unique (s-like orbital).
We denote with $\ket{-}_n$ the ground state of the \textit{n}-th atom, while we use $\ket{+_\alpha}_n$ for the excited states in the three different orientations ($\alpha \in \{1,2,3\}$), which are orthogonal to each other. Thus, we can define generalized orientation-dependent Pauli operators, describing the transition in the respective orientation. 
In particular, we define the generalized $S^z_n = \frac{\hbar}{2} \sigma^z_n$ operator (where $n=1,\dots,N$ is the index of the atom) such that it satisfies the usual relations for spin-$\frac{1}{2}$ systems
\begin{equation}
    S^z \ket{-} = -\frac{\hbar}{2} \ket{-} \,\, , \quad S^z \ket{+_\alpha} = \frac{\hbar}{2} \ket{+_\alpha} \quad \quad \forall \, \alpha \in \{1,2,3\} \, , \label{app:sz}
\end{equation}
where we dropped the atomic index $n$ for notational convenience. Compactly, we can write the previous relations as $S^z \ket{m_\alpha} = \hbar m \ket{m_\alpha}$, where $\ket{m_\alpha} \in \{\ket{-}, \ket{+_\alpha}\}$ and $m \in \{-\frac{1}{2}, \frac{1}{2}\}$ are the respective eigenvalues. Analogously, we define the orientation-dependent raising and lowering operators $S^\pm_\alpha = \hbar \sigma^\pm_\alpha$ which raise or lower the state in the corresponding direction, characterized by the following properties
\begin{equation}
    S^+_\alpha \ket{-} = \hbar \ket{+_\alpha} \,\, , \quad S^+_\alpha \ket{+_\beta} = 0 \,\, , \quad S^-_\alpha \ket{-} = 0 \,\, , \quad S^-_\alpha \ket{+_\beta} = \delta_{\alpha, \beta} \hbar \ket{-} \quad \quad \forall \, \alpha , \beta \in \{1,2,3\} \, , \label{app:spm}
\end{equation}
In the basis $\{\ket{+_1},\ket{+_2},\ket{+_3},\ket{-}\}$, these Pauli operators have the following matrix representation:
\begin{eqnarray} \label{app:sigma_alpha}
    \sigma^-_\alpha & = &
    \begin{pmatrix}
        0 & 0 & 0 & 0 \\
        0 & 0 & 0 & 0 \\
        0 & 0 & 0 & 0 \\
        \delta_{\alpha 1} & \delta_{\alpha 2} & \delta_{\alpha 3} & 0
    \end{pmatrix} \, ,\\
    \sigma^+_\alpha & = & \left( \sigma^-_\alpha \right)^\dag \, , \\
    \sigma^x_\alpha & = & \sigma^-_\alpha + \sigma^+_\alpha \, , \\
    \sigma^y_\alpha & = & i \left( \sigma^-_\alpha - \sigma^+_\alpha \right) \, , \\
    \sigma^z & = &
    \begin{pmatrix}
        1 & 0 & 0 & 0 \\
        0 & 1 & 0 & 0 \\
        0 & 0 & 1 & 0 \\
        0 & 0 & 0 & -1
    \end{pmatrix} \, .
\end{eqnarray}

It can be easily shown that these operators satisfy generalized commutation relations expected from angular-momentum-like operators, such as $\left[ S^z , S^\pm_\alpha \right] = \pm \hbar S^\pm_\alpha$. Using these notions, we can express component-wise the dipole moment operator of the \textit{n}-th atom as 
\begin{equation}
    d_{n,\alpha} = d \sigma^x_{n,\alpha} = d \left( \sigma^-_{n,\alpha} + \sigma^+_{n,\alpha} \right) \, , \label{app:dipole}
\end{equation}
where we considered the dipole moment having the same modulus in the different directions. Hence, we have
\begin{equation}
    {\bf d}_n = \sum_\alpha d_{n,\alpha} {\bf e}_{\alpha} = \sum_\alpha d \left( \sigma^-_{n,\alpha} + \sigma^+_{n,\alpha} \right) {\bf e}_{\alpha} \, . \label{app:dipole_vec}
\end{equation}
Notice that definitions (\ref{app:dipole}) and (\ref{app:dipole_vec}) are consistent with the selection rules prohibiting the transition between states of the same parity.

\section{Generalized Holstein-Primakoff transformations} \label{sec:generalized_HP}

We create generalized Holstein-Primakoff transformations which map the spin operators, defined by Eqs. (\ref{app:sz}) and (\ref{app:spm}), into bosonic operators $b_{\bf k}$. Initially, we will bosonize a two-level system with a threefold degenerated excited state, and successively extend the mapping to a collection of such systems.

To this end, we firstly derive a closed form for the action of $S^\pm_\alpha$ on a generic state $\ket{m_\beta}$. We notice that, by definition, $S^\pm_\alpha \ket{m_\beta} = c^\pm_{\alpha \beta} \ket{m \pm 1_\beta}$. The coefficients $c^\pm_{\alpha \beta}$ are determined by the following relation
\begin{equation}
    | c^\pm_{\alpha \beta} |^2 = \| S^\pm_\alpha \ket{m_\beta} \|^2 = \bra{m_\beta} S^\mp_\alpha S^\pm_\alpha \ket{m_\beta} = \hbar^2 \left[ s \left(s + \delta_{\alpha \beta} \right) - m \left( m \pm \delta_{\alpha \beta} \right) \right] \, , \label{app:c^2}
\end{equation}
where in this case $s = \frac{1}{2}$. Thus, consistently with the Condon-Shortley phase convention, we choose the coefficients to be real and positive
\begin{equation}
    c^\pm_{\alpha \beta} = \hbar \sqrt{s \left(s + \delta_{\alpha \beta} \right) - m \left( m \pm \delta_{\alpha \beta} \right)} = \hbar \sqrt{\left(s \mp m \right) \left( s \pm m + \delta_{\alpha \beta} \right)} \, . \label{app:c_pm}
\end{equation}

Introducing the operator $N = s + S^z /\hbar$, representing the number of excitations in the system, we can relabel the eigenvectors of $S^z$ by noticing that they are clearly also eigenvectors of $N$. Thus, we have $N \ket{n_\alpha} = n \ket{n_\alpha}$, with $n = \frac{1}{2} + m \in \{ 0, 1 \}$. We notice that $n=0$ if the system is in the ground state, while $n=1$ if the system is in its excited state, consistently with the interpretation of the operator $N$. From Eqs.~(\ref{app:sz}) and (\ref{app:c_pm}), the actions of  $S^z$ and $S^\pm_\alpha$ on $\ket{n_\beta}$ in terms of the excitation number are
\begin{eqnarray}
    S^z \ket{n_\beta} & = & \hbar \left( n - \frac{1}{2} \right) \ket{n_\beta} \, , \label{app:Sz_n} \\
    S^+_\alpha \ket{n_\beta} & = & \hbar \sqrt{n+\delta_{\alpha \beta}} \sqrt{1-n} \ket{n+1_\beta} \equiv \hbar \delta_{\alpha \beta} \sqrt{n+ 1} \sqrt{1-n} \ket{n+1_\beta} \, , \label{app:Sp_n} \\
    S^-_\alpha \ket{n_\beta} & = & \hbar \sqrt{1-(n-\delta_{\alpha \beta})} \sqrt{n} \ket{n-1_\beta} \equiv \hbar \delta_{\alpha \beta} \sqrt{1-(n-1)} \sqrt{n} \ket{n-1_\beta} \, , \label{app:Sm_n}
\end{eqnarray}
where in the last steps of Eqs.~(\ref{app:Sp_n}) and (\ref{app:Sm_n}) we exploited the isomorphism between the two members. These relations are indeed consistent with the physical interpretation of $S^\pm_\alpha$, since we expect the operator $S^+_\alpha (S^-_\alpha)$ to be able to increase (decrease) the quantum number only in the $\alpha$-direction.

We can now define bosonic operators for the different orientations $b_\alpha$, which satisfy the usual properties
\begin{eqnarray} 
    b_\alpha \ket{n_\beta} & = & \delta_{\alpha \beta} \sqrt{n} \ket{n-1_\beta} \, , \label{app:annih} \\
    b^\dag_\alpha \ket{n_\beta} & = & \delta_{\alpha \beta} \sqrt{n+1} \ket{n+1_\beta} \, , \label{app:creat} \\
    b^\dag_\alpha b_\alpha \ket{n_\beta} & = & \delta_{\alpha \beta} n \ket{n_\beta} \, , \label{app:num} \\
    \left[ b_\alpha, b^\dag_\beta \right] & = & \delta_{\alpha \beta} \, . \label{app:comm}
\end{eqnarray}
Therefore, using the previous relations (\ref{app:Sz_n}-\ref{app:Sm_n}), we can establish the generalized Holstein-Primakoff transformations for a single two-level system, which relate the spin to the bosonic operators, as
\begin{eqnarray}
    S^+_\alpha & = & \hbar b^\dag_\alpha \sqrt{1-\sum_\beta b^\dag_\beta b_\beta} \, , \label{app:S^+} \\
    S^-_\alpha & = & \hbar \sqrt{1-\sum_\beta b^\dag_\beta b_\beta} \;  b_\alpha \, , \label{app:S^-} \\
    S^z & = & \hbar \left( \sum_\beta b^\dag_\beta b_\beta - \frac{1}{2} \right) \, . \label{app:S^z}
\end{eqnarray}
In the limit of low excitations in the system (corresponding to a low average excitation per site), we can expand the radicals in Eqs.~(\ref{app:S^+}) and (\ref{app:S^-}) and retain only the lowest power term in each expression, which yields to
\begin{equation}
    S^{-(+)}_\alpha \approx \hbar b^{(\dag)}_\alpha \, . \label{app:S_pm_termlim}
\end{equation}

Finally, we consider a collection of $N$ spin-$\frac{1}{2}$ identical systems and construct collective operators in the three-dimensional {\bf k}-space. In particular, labeling with $b_{n, \alpha}$ the bosonic operator associated with the $n$-th site, we define the collective bosonic operators $b_{{\bf k},\alpha}$ as
\begin{equation}
    b_{{\bf k},\alpha} = \frac{1}{\sqrt{N}} \sum_n e^{-i {\bf k} \cdot {\bf R}_n} b_{n, \alpha} \, . \label{app:b_k}
\end{equation}
It can be readily verified that these operators obey the commutation relations $\left[ b_{{\bf k},\alpha}, b_{{\bf k}',\beta} \right] = \delta_{{\bf kk}'} \delta_{\alpha \beta}$. In the derivation of the Hamiltonians, we will make use of the relation $\sum_n b^\dag_{n, \alpha} b_{n, \alpha} = \sum_{\bf k} b^\dag_{{\bf k},\alpha} b_{{\bf k},\alpha}$, which can be easily demonstrated. 
Furthermore, using Eqs.~(\ref{app:S_pm_termlim}) and (\ref{app:b_k}), we derive the relations (valid in the low-excitation regime)
\begin{equation}
    \sum^{N}_{n=1} e^{- (+) i {\bf k} \cdot {\bf R}_n} S^{-(+)}_{n, \alpha} = \hbar \sqrt{N} b^{(\dag)}_{{\bf k},\alpha} \, . \label{app:final_rel}
\end{equation}

\section{Dipole-dipole interactions} \label{sec:dipolar_int}

The electrostatic dipole-dipole interaction between two localized charges is expressed by
\begin{equation}
    V_{\rm dip} \left( {\bf R}_1 , {\bf R}_2 \right) = \frac{1}{4 \pi \epsilon_0 \epsilon_m} \frac{{\bf d}_1 \cdot {\bf d}_2 - 3 ({\bf d}_1 \cdot \hat{\bf r}_{12})({\bf d}_2 \cdot \hat{\bf r}_{12})}{r_{12}^3} \, . \label{app:V_dip}
\end{equation}
Therefore, the dipole-dipole Hamiltonian contribution for a three-dimensional system of ordered particles is given (as reported in the main paper) by
\begin{equation}
    H_{\rm dip} = \frac{1}{2} \sum_{n \neq m} V_{dip} \left( {\bf R}_n , {\bf R}_m \right) = \frac{1}{8 \pi \epsilon_0 \epsilon_m} \sum_{n \neq m} \frac{{\bf d}_n \cdot {\bf d}_m - 3 ({\bf d}_n \cdot \hat{\bf r}_{nm})({\bf d}_m \cdot \hat{\bf r}_{nm})}{r_{nm}^3} \, . \label{app:H_dip_dd}
\end{equation}
We can now perform the two-level approximation (\ref{app:dipole_vec}) and successively proceed to the bosonization of the system in the thermodynamic limit through the use of the generalized Holstein-Primakoff transformations (see Sec. \ref{sec:generalized_HP}). Thus, in the long-wavelength approximation, we can express the matter field in terms of collective bosonic operators $b_{{\bf k}, \alpha}$. Therefore, after the bosonization, the dipole-dipole term in \eqref{app:H_dip_dd} is expressed as
\begin{equation}
    H_{\rm dip} = \sum_{\alpha, \beta, {\bf k}} F_{{\bf k}, \alpha, \beta} \left( b_{{\bf k}, \alpha} + b^\dag_{-{\bf k}, \alpha} \right) \left( b_{-{\bf k}, \beta} + b^\dag_{{\bf k}, \beta} \right) \, , \label{app:H_dip}
\end{equation}
where we have defined structure-dependent factor $F_{{\bf k}, \alpha, \beta}$ as
\begin{equation}
    F_{{\bf k}, \alpha, \beta} = \frac{d^2}{8 \pi \epsilon_m}  \sum_{l \neq 0} \frac{\cos{{\bf k} \cdot {\bf r}_l}}{r_l^3} \left( \delta_{\alpha, \beta} - 3 (\hat{{\bf e}}_\alpha \cdot \hat{\bf r}_l) (\hat{{\bf e}}_\beta \cdot \hat{\bf r}_l) \right) \, , \label{app:factor_F}
\end{equation}
where we used the translational symmetry of the system. ${\bf r}_l \equiv {\bf r}_{0l}$ is the distance of the $l$-th site from the origin, which is the only point excluded in the summation ($l \neq 0$). Such factor can be approximated, in the long-wavelength limit (corresponding to the neighborhood of the $\Gamma$ point in the crystal), as \cite{Cohen_Keffer1955}
\begin{equation}
    \sum_{l \neq 0} \frac{\cos{{\bf k} \cdot {\bf r}_l}}{r_l^3} \left( \delta_{\alpha, \beta} - 3 (\hat{{\bf e}}_\alpha \cdot \hat{\bf r}_l) (\hat{{\bf e}}_\beta \cdot \hat{\bf r}_l) \right) \approx \frac{4 \pi}{3 v} \rho \left( 3 (\hat{\bf k} \cdot \hat{{\bf e}}_\alpha) (\hat{\bf k} \cdot \hat{{\bf e}}_\beta) - \delta_{\alpha \beta} \right) \, ,
\end{equation}
where the factor $v$ depends on the lattice structure and it is equal to $v = 1$ for sc, $v = 2^{-1/2}$ for fcc and $v = 4 \cdot 3^{-3/2}$ for bcc lattices.
Thus, we can define the factor $f_{{\bf k}, \alpha, \beta} \approx \left( 3 (\hat{\bf k} \cdot \hat{{\bf e}}_\alpha) (\hat{\bf k} \cdot \hat{{\bf e}}_\alpha) - \delta_{\alpha \beta}\right) / 3$, which directly follows from the form of $F_{{\bf k}, \alpha, \beta}$, as in the main text. Hence, we have for a three-dimensional lattice in the long-wavelength approximation
\begin{equation}
    F_{{\bf k}, \alpha, \beta} = \frac{d^2 \rho}{2 \epsilon_0 \epsilon_m} \frac{3 (\hat{\bf k} \cdot \hat{{\bf e}}_\alpha) (\hat{\bf k} \cdot \hat{{\bf e}}_\beta) - \delta_{\alpha \beta}}{3} = \hbar \chi^2 \omega_0 f_{{\bf k}, \alpha, \beta} = \hbar \eta^2 \omega_0 f_{{\bf k}, \alpha, \beta} \, ,
    \label{app:factor_F_bulk}
\end{equation}
where we used the relation $\chi^2 \omega_0 = d^2 \rho / 2 \hbar \epsilon_0 \epsilon_m$ and the equality $\chi = \eta$, \textit{valid only in the three-dimensional case}.

On the other hand, it is instructive to repeat the procedure for a square (bi-dimensional) lattice. The main difference from the three-dimensional case is that the system has not any discrete translational invariance along the $z$-axis, given that the charges are ordered in a planar structure on the $xy$ plane. Thus, the mode decomposition (and thus the matter bosonic operators $b_{{\bf k}_\|, \alpha}$) only depends on ${\bf k}_\|$. Hence, for incidence orthogonal to the plane surface, Eq.~(\ref{app:H_dip}) becomes
\begin{equation}
    H_{\rm dip} = \sum_{\alpha, \beta} F_{\alpha, \beta} \left( b_{\alpha} + b^\dag_{\alpha} \right) \left( b_{\beta} + b^\dag_{\beta} \right) \, , \label{app:H_dip_layer}
\end{equation}
where the factor $F_{\alpha, \beta}$ reduces to
\begin{equation}
    F_{\alpha, \beta} = \frac{d^2}{8 \pi \epsilon_0 \epsilon_m}  \sum_{l \neq 0} \frac{1}{r_l^3} \left( \delta_{\alpha, \beta} - 3 (\hat{{\bf e}}_\alpha \cdot \hat{\bf r}_l) (\hat{{\bf e}}_\beta \cdot \hat{\bf r}_l) \right) \, . \label{app:factor_F_layer}
\end{equation}
In particular, this expression is independent on the wavevector. Thus, it doesn't rely on the limit of long-wavelength to be evaluated. In fact, being $a$ the lattice constant, Eq.~(\ref{app:factor_F_layer}) can be evaluated as 
\begin{equation}
    F_{\alpha, \beta} = \frac{d^2 \mu}{\epsilon_0 \epsilon_m a^3} \frac{3 (\hat{\bf z} \cdot \hat{{\bf e}}_\alpha) (\hat{\bf z} \cdot \hat{{\bf e}}_\beta) - \delta_{\alpha \beta} }{3} = \hbar \chi^2 \omega_0 f_{{\bf z}, \alpha, \beta} \, , \label{app:factor_F_layer_final}
\end{equation}
where we defined for the 2D lattice
\begin{eqnarray}
    & \chi = \sqrt{\frac{d^2 \mu}{\hbar \epsilon_0 \epsilon_m a^3 \omega_0}} \, , \\
    & \mu = \frac{3}{4 \pi} \left[ \sum_{n_x > 0} n_x^{-3} +  \sum_{n_x, n_y > 0} \left( n_x^2 + n_y^2 \right)^{-3/2} \right] \approx \frac{6.78}{4 \pi} \, .
\end{eqnarray}

As pointed out in the main text, the crucial difference between the two-dimensional and three-dimensional cases is the dependence of the light-matter coupling on the lattice constant $a$. Firstly, we notice that the factor $F_{{\bf k}, \alpha, \beta} \propto \chi^2$ has the same dependence on the lattice constant $a$ in both cases, i.e. $\chi^2 \propto a^{-3}$. On the other hand, in the 3D case we can associate the light-matter coupling to the volumetric density $rho$, which in turn is related to the lattice constant through $\rho \propto a^{-3}$. In contrast, in the 2D lattice we have a light-matter constant dependent on a superficial density $\sigma \propto a^{-2}$ (it also depend on the cavity length $L$, which on the contrary doesn't play any role in the dipole-dipole interaction term as it is to be expected). Therefore, for such a system, it can be clearly understood that the eventual softening of the transverse matter frequency is purely due to the dipolar interactions, independently on the coupling with the radiation field, which depends also on other parameters.

\section{Derivation of the Hamiltonians} \label{sec:derivation_H} 

In this section we present the full derivation of the Hamiltonian in the 3D and 2D lattice. In the next Section \ref{sec:disp_rel}, we proceed to introduce the diagonalization procedure for the full Hamiltonians and compare the results obtained with the results presented in the main paper, which instead follow a two-step Bogoliubov transformation (firstly on the matter subsystem and successively considering the interaction with the radiation field), thus demonstrating the equivalence of the two procedures.

\subsection{3D lattice}

Let us consider a system composed of a three-dimensional bulk of atoms interacting with a quantized electromagnetic field. As usual, the vector potential $\bf A(r)$ is quantized by introducing radiation bosonic operators $a_{{\bf k}, \lambda}$ for each mode $\bf k$ and polarization $\lambda$, leading to the plane-wave decomposition
\begin{equation}
    {\bf A}({\bf r}) = \sum_{\lambda} \sum_{\bf k} {\cal E}_k e^{i{\bf k} \cdot {\bf r} } a_{{\bf k}, \lambda}  \hat{{\bf e}}_\lambda + {\rm H.c.} \, , \label{app:vec_pot_bulk}
\end{equation}
where ${\cal E}_k = \sqrt{{\hbar}/{2 \epsilon_m \omega_k V}}$, with $V$ being the quantization volume, and $\hat{\bf e}_\lambda$ ($\lambda=1,2$) are the two polarization unit vectors orthogonal to $\hat{\bf k}$. Moreover, as usual, the Hamiltonian of the free electromagnetic field is given by
\begin{equation}
    H_{\rm ph} = \sum_{\lambda, {\bf k}} \hbar \omega_k a^\dag_{{\bf k}, \lambda} a_{{\bf k}, \lambda} \, . \label{app:H_ph}
\end{equation}
On the other hand, the atoms in this system are modeled as localized charges in the sites ${\bf R}_n$ of a lattice. In the long-wavelength approximation, the total polarization density can be expressed as ${\bf P(r)} = \sum_n {\bf d}_n \delta ({\bf r} - {\bf R}_n)$, where ${\bf d}_n$ is the total electric dipole of the \textit{n}-th atom, thus assuming non-overlapping dipoles for different atoms. In the two-level approximation (see Sec.~\ref{sec:two_lev}), the Hamiltonian of $N$ non-interacting atoms can be written as
\begin{equation}
    H_{\rm A} = \sum_n \frac{\hbar \omega_0}{2} \sigma^z_n \, . \label{app:H_A}
\end{equation}
In order to derive the full light-matter Hamiltonian in the multipolar gauge \cite{Cohen_book}, we apply to the Hamiltonian of the free electromagnetic field (\ref{app:H_ph}) the unitary transformation $U$ defined by \cite{Garziano2020}
\begin{equation}
    U = \exp \left( \frac{i}{\hbar} \int {\bf A(r) \cdot P(r)} d^3 {\bf r} \right) \, . \label{app:U_def}
\end{equation}
Inserting the previous definitions for the vector potential and polarization in Eq.~(\ref{app:U_def}), we obtain the transformation
\begin{equation}
    U = \exp \left[ \frac{i}{\hbar} \sum_{\lambda, {\bf k}, n} {\cal E}_{\bf k} \left( e^{i{\bf k} \cdot {\bf R}_n } a_{{\bf k}, \lambda}+ e^{-i{\bf k} \cdot {\bf R}_n } a^\dag_{{\bf k}, \lambda} \right) \hat{{\bf e}}_\lambda \cdot {\bf d}_n \right] \, . \label{app:U}
\end{equation}
A direct application of the transformation (\ref{app:U}) to the photonic Hamiltonian (\ref{app:H_ph}) yields
\begin{equation}
    U^\dag H_{\rm ph} U = \sum_{\lambda, {\bf k}} \hbar \omega_k \left[ a^\dag_{{\bf k}, \lambda} a_{{\bf k}, \lambda} 
    - \frac{i}{\hbar} \sum_n {\cal E}_{\bf k} \left( e^{i{\bf k} \cdot {\bf R}_n } a_{{\bf k}, \lambda} - e^{-i{\bf k} \cdot {\bf R}_n } a^\dag_{{\bf k}, \lambda} \right) \hat{{\bf e}}_\lambda \cdot {\bf d}_n 
    + \frac{1}{\hbar^2} \sum_{n,m} {\cal E}_{\bf k}^2 \, e^{i {\bf k} \cdot ({\bf R}_n - {\bf R}_m)} \, \hat{{\bf e}}_\lambda \cdot {\bf d}_n \, \hat{{\bf e}}_\lambda \cdot {\bf d}_m \right]. \label{app:H_ph_bulk_transf}
\end{equation}

We can now bosonize the system through the use of generalized Holstein-Primakoff transformations (see Sec.~\ref{sec:generalized_HP}), leading to the atomic and transformed photonic Hamiltonians, respectively
\begin{eqnarray}
    H_{\rm A} & = & \hbar \omega_0 \sum_{\alpha, {\bf k}} b^\dag_{{\bf k}, \alpha} b^{}_{{\bf k}, \alpha} \, , \\
    U^\dag H_{\rm ph} U & = & \hbar \sum_{\lambda, {\bf k}} \omega_k a^\dag_{{\bf k}, \lambda} a^{}_{{\bf k}, \lambda} 
    - i \hbar \sum_{\alpha, \lambda, {\bf k}} g_k \omega_k \left( a^{}_{{\bf k}, \lambda} - a^\dag_{-{\bf k}, \lambda} \right) \left( b^{}_{-{\bf k}, \alpha} + b^\dag_{{\bf k}, \alpha} \right) e_{\lambda_{\alpha}} \nonumber \\
    & & + \hbar \sum_{\alpha, \beta, \lambda, {\bf k}} g^2_k \omega_k\left( b^{}_{-{\bf k}, \alpha} + b^\dag_{{\bf k}, \alpha} \right) \left( b^{}_{{\bf k}, \beta} + b^\dag_{-{\bf k}, \beta} \right) e_{\lambda_{\alpha}} e_{\lambda_{\beta}} \, ,
\end{eqnarray}
where $g_k = {\cal E}_{\bf k} d \sqrt{N} / \hbar = \sqrt{d^2 N / 2 \hbar \epsilon_m V \omega_k}$ and the $\hat{{\bf e}}_\alpha$ are a generic set of orthonormal basis vectors used as a basis for the decomposition of the dipole moments. Furthermore, we indicated for notation convenience $e_{\lambda_{\alpha}} \equiv \hat{{\bf e}}_\lambda \cdot \hat{{\bf e}}_\alpha$. 
The additional term describing the dipole-dipole interactions can be expressed (see Sec. \ref{sec:dipolar_int}) in the thermodynamic limit as
\begin{equation}
    H_{\rm dip} = \hbar \sum_{\alpha, \beta, {\bf k}} \chi^2 \omega_0 f_{{\bf k}, \alpha, \beta} \left( b^{}_{-{\bf k}, \alpha} + b^\dag_{{\bf k}, \alpha} \right) \left( b^{}_{{\bf k}, \beta} + b^\dag_{-{\bf k}, \beta} \right) \, .
\end{equation}
Hence, the total Hamiltonian in the multipolar gauge is given by
\begin{eqnarray}
    H = & & \hbar \sum_{\lambda, {\bf k}} \omega_k a^\dag_{{\bf k}, \lambda} a^{}_{{\bf k}, \lambda} + \hbar \omega_0 \sum_{\alpha, {\bf k}} b^\dag_{{\bf k}, \alpha} b^{}_{{\bf k}, \alpha} \nonumber \\
    & - & i \hbar \sum_{\alpha, \lambda, {\bf k}} g_k \omega_k \left( a_{{\bf k}, \lambda} - a^\dag_{-{\bf k}, \lambda} \right) \left( b^{}_{-{\bf k}, \alpha} + b^\dag_{{\bf k}, \alpha} \right) e_{\lambda_{\alpha}} \nonumber \\
    & + & \hbar \sum_{\alpha, \beta, \lambda, {\bf k}} g^2_k \omega_k\left( b^{}_{-{\bf k}, \alpha} + b^\dag_{{\bf k}, \alpha} \right) \left( b^{}_{{\bf k}, \beta} + b^\dag_{-{\bf k}, \beta} \right) e_{\lambda_{\alpha}} e_{\lambda_{\beta}} \nonumber \\
    & + & \hbar \sum_{\alpha, \beta, {\bf k}} \chi^2 \omega_0 f_{{\bf k}, \alpha, \beta} \left( b^{}_{-{\bf k}, \alpha} + b^\dag_{{\bf k}, \alpha} \right) \left( b^{}_{{\bf k}, \beta} + b^\dag_{-{\bf k}, \beta} \right) \, . \label{app:H_M_bulk_full}
\end{eqnarray}
The first two terms in Eq.~(\ref{app:H_M_bulk_full}) represent the free radiation and matter fields, the third and the fourth are the transverse interactions between light and matter, while the last one represent the dipole-dipole electrostatic interactions, respectively.

\subsection{2D lattice}

Let us now consider a system composed of a two-dimensional layer of atoms, identified as the $xy$ plane, embedded in an ideal cavity. This planar configuration naturally induces a decomposition of the vector potential as
\begin{equation}
    A ({\bf r}) = \sum_{\lambda, {\bf k}_\|} \sum_{k_z > 0} {\cal E}_k e^{i{\bf k}_\| \cdot {\bf r_\|}}  \left( e^{i k_z z} a_{l,{\bf k_\|}, k_z, \lambda} + e^{-i k_z z} a_{r,{\bf k_\|}, k_z, \lambda} \right)  \hat{\bf e}_\lambda  + {\rm H.c.} \, ,
    \label{app:vec_pot_pos_layer}
\end{equation}
where ${\bf k}_\|$ is the in-plane discrete component of the wave vector, being $S$ the corresponding quantization surface, while $k_z$ is its orthogonal component quantized by the length of the cavity $L$, as already stated in the main text. In this expression, ${\cal E}_k = \sqrt{ \hbar / 2 \epsilon_m \omega_k S L}$ and $\omega_k = v \sqrt{{\bf k_\|}^2 + k_z^2}$. Moreover, we introduced the left and right creation operators $a_{l(r),{\bf k_\|}, k_z, \lambda} \equiv a_{{\bf k_\|}, \pm k_z, \lambda}$, where the index $l\,(r)$ is associated with the $+(-)$ sign. If we consider the case of normal incidence $\bf k_\| =0$, and thus ${\bf k} = k_z {\bf \hat{z}}$, Eq. (\ref{app:vec_pot_pos_layer}) considerably simplifies given that the two polarization vectors $\hat{{\bf e}}_\lambda$ now lie in the $xy$ plane and the bosonic operators become independent on ${\bf k}_\|$, i.e. $a_{l(r), k_z, \lambda} \equiv a_{l(r),{\bf k_\| = 0}, k_z, \lambda}$. We can define the even and odd radiation modes operators as $a_{e(o), k_z, \lambda} =  \left( a_{l, k_z, \lambda} \pm a_{r, k_z, \lambda} \right)/\sqrt{2}$.
Therefore, the Hamiltonian of the free electromagnetic field can be written as
\begin{equation}
    H_{\rm ph} = \sum_\lambda \sum_{j=e,o} \sum_{k_z > 0} \hbar \omega_{k_z} a^\dag_{j, k_z, \lambda} a_{j, k_z, \lambda} \, . \label{app:H_ph_layer}
\end{equation}

For the matter subsystem we follow an analogous procedure to the 3D lattice of dipoles. We firstly perform the two-level approximation, taking into account the dipole orientations, and successively construct two-dimensional collective bosonic operators $b_{\bf k_\|}$. 
Considering only radiation modes with wavevectors orthogonal to the 2D lattice surface (normal incidence), implying ${\bf k}_\| = 0$, the atomic Hamiltonian reads
\begin{equation}
    H_{\rm A} = \hbar \omega_0 \sum_{\alpha} b^\dag_{\alpha} b_{\alpha} \, .
\end{equation}

We now calculate the transformation $U$, analogous to \eqref{app:U_def}, for such a system, yielding to
\begin{equation}
    U = \exp \left[ \frac{i}{\hbar} \sum_{\lambda, n} \sum_{k_z>0} \sqrt{2} {\cal E}_{k_z} \left( a_{e, k_z, \lambda} + a^\dag_{e, k_z, \lambda} \right) \hat{{\bf e}}_\lambda \cdot {\bf d}_n \right] \, ,
\end{equation}
where, again, normal incidence has been considered. Applying this transformation to the photonic Hamiltonian \eqref{app:H_ph_layer}, we obtain
\begin{equation}
    U^\dag H_{\rm ph} U = \sum_{\lambda, k_z > 0} \hbar \omega_{k_z} \left[ \sum_{j=e,o} a^\dag_{j, k_z, \lambda} a_{j, k_z, \lambda}
    - \frac{i}{\hbar} \sum_n \sqrt{2} {\cal E}_{k_z} \left( a_{e, k_z, \lambda} - a^\dag_{e, k_z, \lambda} \right) \hat{{\bf e}}_\lambda \cdot {\bf d}_n 
    + \frac{1}{\hbar^2} \sum_{n,m} 2 {\cal E}_{k_z}^2 \, \hat{{\bf e}}_\lambda \cdot {\bf d}_n \, \hat{{\bf e}}_\lambda \cdot {\bf d}_m \right] \, . \label{app:H_ph_layer_transf}
\end{equation}

Therefore, after the bosonization in the thermodynamic limit, the full light-matter Hamiltonian, considering the addition of the dipole-dipole interactions in a planar layer \eqref{app:factor_F_layer_final}, is given by
\begin{eqnarray}
    H = & & \hbar \sum_{\lambda, k_z > 0} \omega_{k_z} \left( a^\dag_{e, k_z, \lambda} a_{e, k_z, \lambda} + a^\dag_{o, k_z, \lambda} a_{o, k_z, \lambda} \right) + \hbar \omega_0 \sum_{\alpha} b^\dag_{\alpha} b^{}_{\alpha} \nonumber \\
    & - & i \hbar \sum_{\alpha, \lambda, k_z > 0} g_{k_z} \omega_{k_z} \left( a_{e, k_z, \lambda} - a^\dag_{e, k_z, \lambda} \right) \left( b_{\alpha} + b^\dag_{\alpha} \right) e_{\lambda_{\alpha}} \nonumber \\
    & + & \hbar \sum_{\alpha, \beta, \lambda, k_z > 0} g^2_{k_z} \omega_{k_z} \left( b^{}_{\alpha} + b^\dag_{\alpha} \right) \left( b^{}_{\beta} + b^\dag_{\beta} \right) e_{\lambda_{\alpha}} e_{\lambda_{\beta}} \nonumber \\
    & + & \hbar \sum_{\alpha, \beta} \chi^2 \omega_0 f_{{\bf z}, \alpha, \beta} \left( b^{}_{\alpha} + b^\dag_{\alpha} \right) \left( b^{}_{\beta} + b^\dag_{\beta} \right) \, , \label{app:H_M_layer_full}
\end{eqnarray}
where $g_k = \sqrt{d^2 N / \hbar \epsilon_m S L \omega_{k_z}}$. As a final note, we remark that the Coulomb gauge Hamiltonians can be derived by applying the minimal coupling replacement to the corresponding matter Hamiltonians instead, leading, nonetheless, to exactly the same dispersion relations as in the multipolar gauge.

\section{Calculation of the dispersion relations} \label{sec:disp_rel}

The dispersion relations can be calculated through the use of a Hopfield-Bogoliubov diagonalization of the total system Hamiltonian, e.g. 3D lattice Hamiltonian (\ref{app:H_M_bulk_full}), or equivalently, by a two-step diagonalization as presented in the main paper. In this section we will follow the first procedure, and verify the equivalence of the results.

Firstly, let us evaluate the following commutators
\begin{eqnarray}
    \left[ a_{{\bf k}, \lambda} , H \right] & = & \hbar \omega_k a_{{\bf k}, \lambda} + i \hbar g_k \omega_k \sum_{\alpha} \left( b^\dag_{-{\bf k}, \alpha} + b_{{\bf k}, \alpha} \right) e_{\lambda_{\alpha}} \\
    \left[ a^\dag_{{\bf k}, \lambda} , H \right] & = & -\hbar \omega_k a^\dag_{{\bf k}, \lambda} + i \hbar g_k \omega_k \sum_{\alpha} \left( b^\dag_{{\bf k}, \alpha} + b_{-{\bf k}, \alpha} \right) e_{\lambda_{\alpha}} \\
    \left[ b_{{\bf k}, \alpha} , H \right] & = & \hbar \omega_0 b_{{\bf k}, \alpha} + i \hbar g_k \omega_k \sum_{\lambda} \left( a^\dag_{-{\bf k}, \lambda} - a_{{\bf k}, \lambda} \right) e_{\lambda_{\alpha}} + 2 \hbar g^2_k \omega_k \sum_{\beta} \left[ \sum_{\lambda} e_{\lambda_{\beta}} e_{\lambda_{\alpha}} + f_{{\bf k}, \alpha, \beta}  \right] \left( b^\dag_{-{\bf k}, \beta} + b_{{\bf k}, \beta} \right) \\
    \left[ b^\dag_{{\bf k}, \alpha} , H \right] & = & - \hbar \omega_0 b^\dag_{{\bf k}, \alpha} - i \hbar g_k \omega_k \sum_{\lambda} \left( a^\dag_{{\bf k}, \lambda} - a_{-{\bf k}, \lambda} \right) e_{\lambda_{\alpha}} - 2 \hbar g^2_k \omega_k \sum_{\beta} \left[ \sum_{\lambda} e_{\lambda_{\beta}} e_{\lambda_{\alpha}} + f_{{\bf k}, \alpha, \beta}  \right] \left( b^\dag_{{\bf k}, \beta} + b_{-{\bf k}, \beta} \right).
\end{eqnarray}

Thus, writing the Heisenberg equations and transforming to the frequency-domain Fourier space, we have
\begin{eqnarray}
    \hbar \left( \Omega - \omega_k \right) \mathcal{A}_{{\bf k}, \lambda} & = & i \hbar g_k \omega_k \sum_{\alpha} \left( \mathcal{B}^\dag_{-{\bf k}, \alpha} + \mathcal{B}_{{\bf k}, \alpha} \right) e_{\lambda_{\alpha}} \\
    \hbar \left( \Omega + \omega_k \right) \mathcal{A}^\dag_{-{\bf k}, \lambda} & = & i \hbar g_k \omega_k \sum_{\alpha} \left( \mathcal{B}^\dag_{-{\bf k}, \alpha} + \mathcal{B}_{{\bf k}, \alpha} \right) e_{\lambda_{\alpha}} \\
    \hbar \left( \Omega - \omega_0 \right) \mathcal{B}_{{\bf k}, \alpha} & = & i \hbar g_k \omega_k \sum_{\lambda} \left( \mathcal{A}^\dag_{-{\bf k}, \lambda} - \mathcal{A}_{{\bf k}, \lambda} \right) e_{\lambda_{\alpha}} + 2 \hbar g^2_k \omega_k \sum_{\beta} \left[ \sum_{\lambda} e_{\lambda_{\beta}} e_{\lambda_{\alpha}} + f_{{\bf k}, \alpha, \beta}  \right] \left( \mathcal{B}^\dag_{-{\bf k}, \beta} + \mathcal{B}_{{\bf k}, \beta} \right) \\
    \hbar \left( \Omega + \omega_0 \right) \mathcal{B}^\dag_{-{\bf k}, \alpha} & = & - i \hbar g_k \omega_k \sum_{\lambda} \left( \mathcal{A}^\dag_{-{\bf k}, \lambda} - \mathcal{A}_{{\bf k}, \lambda} \right) e_{\lambda_{\alpha}} - 2 \hbar g^2_k \omega_k \sum_{\beta} \left[ \sum_{\lambda} e_{\lambda_{\beta}} e_{\lambda_{\alpha}} + f_{{\bf k}, \alpha, \beta}  \right] \left( \mathcal{B}^\dag_{-{\bf k}, \beta} + \mathcal{B}_{{\bf k}, \beta} \right),
\end{eqnarray}
where $\mathcal{A}_{{\bf k}, \lambda} (\Omega)$ and $\mathcal{B}_{{\bf k}, \lambda} (\Omega)$ are the Fourier transforms of the operators $a_{{\bf k}, \lambda} (t)$ and $b_{{\bf k}, \lambda} (t)$, respectively (the dependencies on time or frequency are omitted for simplicity).

\begin{figure}[b]
    \centering
    \includegraphics[width=0.6\linewidth]{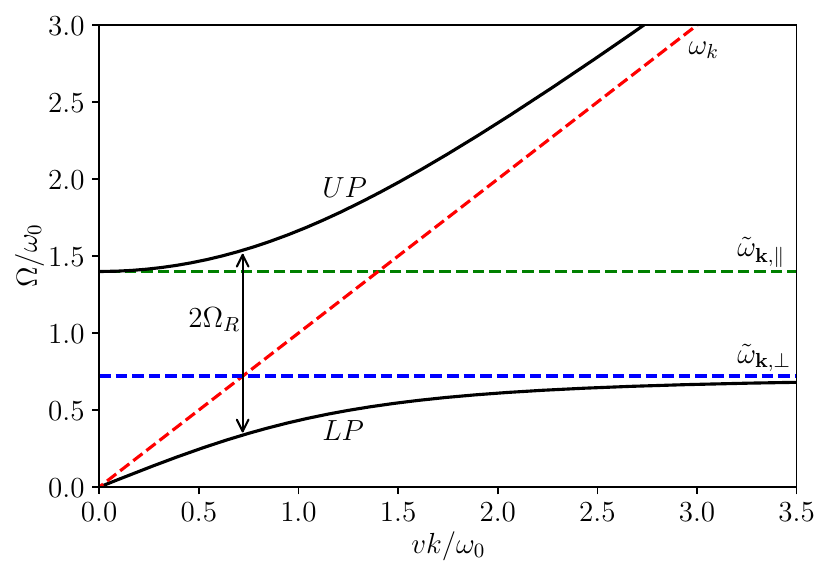}
    \caption{Dispersion curves of a renormalized Hopfield-like model, calculated with $\eta = 0.6$.}
    \label{fig:example_disp_curv}
\end{figure}
Solving this system we can derive self-consistent equations for the transverse and longitudinal sectors. We consider here for simplicity the dipole basis used for the decomposition ${\bf e}_\alpha$ coincident with the basis induced by radiation field and successively select the transverse and longitudinal components, respectively, thus obtaining
\begin{eqnarray}
    \frac{\omega^2_{\bf k}}{\Omega^2_\perp}  & = & 1 + \frac{4 g^2_k \omega_k \omega_0 }{\omega^2_0 + 4 g^2_k \omega_k \omega_0 f^\perp_{\bf k} - \Omega^2_\perp} \, , \label{app:eq_transv_bulk} \\
    \Omega^2_\| & = & \omega^2_0 + 4 g^2_k \omega_k \omega_0 f^\|_{\bf k} \, , \label{app:eq_long_bulk}
\end{eqnarray}
where $f^\perp_{\bf k}$ and $f^\|_{\bf k}$ represent the transverse and longitudinal components of $f_{{\bf k}, \alpha, \beta}$, respectively. For isotropic systems, in the long-wavelength approximation $f^\perp_{\bf k} = -1/3$ and $f^\|_{\bf k} = 2/3$. 
If we introduce a renormalized matter frequency defined, as in the main paper, by $\Tilde{\omega}_{{\bf k}, \alpha} = \omega_0 \sqrt{1 + 4 \eta^2 f_{{\bf k}, \alpha, \alpha}}$, the dispersion relation for the transverse sector (\ref{app:eq_transv_bulk}) becomes equivalent to the one derived by an Hopfield model with renormalized factors
\begin{equation}
    \frac{\omega^2_{\bf k}}{\Omega^2_\perp} = 1 + \frac{4 g'^2_k \omega_k \Tilde{\omega}^\perp_{\bf k}}{\Tilde{\omega}_{\bf k}^{\perp^2} - \Omega^2_\perp} \, ,
\end{equation}
where $g'_k = g_k \sqrt{\omega_0 / \Tilde{\omega}^\perp_{\bf k}}$ is the normalized coupling and $\Tilde{\omega}^\perp_{\bf k} = \omega_0 \sqrt{1 + 4 \eta^2 f^\perp_{\bf k}}$ is the normalized transverse frequency. 
Introducing the coupling constant $\eta = \sqrt{d^2 N / 2 \hbar \epsilon_m V \omega_0}$, the previous dispersion relation can be written as
\begin{equation}
    \frac{\omega^2_{\bf k}}{\Omega^2_\perp} = 1 + \frac{4 \eta^2 \omega^2_0}{\omega^2_0 + 4 \eta^2 \omega^2_0 f^\perp_{\bf k} - \Omega^2_\perp} = 1 + \frac{4 \eta'^2 \Tilde{\omega}_{\bf k}^{\perp^2}}{\Tilde{\omega}_{\bf k}^{\perp^2} - \Omega^2_\perp} \, , \label{app:disp_rel_pre}
\end{equation}
where we defined the renormalized coupling $\eta' = \eta \, \omega_0 / \Tilde{\omega}^\perp_{\bf k}$ in order to reduce the dispersion relation to an Hopfield-like one. Thus we demonstrated the equivalence between this diagonalization procedure of the full Hamiltonian and the two-step diagonalization introduced in the main paper. 
The new coupling constant $\eta$ is particularly useful in the evaluation of the experimental results, since it corresponds to the ratio between the Rabi frequency $\Omega_R$ (which in turn is half the splitting between the level anticrossing) and the matter frequency, i.e. $\eta = \Omega_R / \omega_0$. Analogously, the normalized coupling constant $\eta'$ is related to the normalized matter frequency through $\eta' = \Omega_R / \Tilde{\omega}^\perp_{\bf k}$. In Fig. \ref{fig:example_disp_curv} represents an example of typical dispersion curves, in which all the main physical dispersion are illustrated.

\section{Dispersion relations above phase transition} \label{sec:aboveSPT}

In this section we derive the dispersion relations for Hamiltonian (\ref{app:H_M_bulk_full}) above the phase transition. Such an Hamiltonian has to be modified accordingly for coupling strength greater than the critical value $\eta_c$, given the predicted macroscopic occupation of the fields in this new phase. In order to correctly describe this phenomena, we shift the the bosonic radiation and matter mode operators as \cite{Brandes03}
\begin{eqnarray}
    a_{{\bf k}, \lambda} = \Tilde{a}_{{\bf k}, \lambda} + i A_{{\bf k}, \lambda} \, , \\
    b_{{\bf k}, \alpha} = \Tilde{b}_{{\bf k}, \alpha} - B_{{\bf k}, \alpha} \, ,
\end{eqnarray}
where the parameters $A_{{\bf k}, \lambda}$ and $B_{{\bf k}, \alpha}$ are linked to the macroscopic mean field mode occupation.
Thus, we expect them to be zero in the normal phase and of order $O(\sqrt{N})$ above the phase transition. 
After substitute these relations into Eq.~(\ref{app:H_M_bulk_full}), expanding the square root contribution in the Holstein-Primakoff procedure and retaining only the terms up to second order, the resulting superradiant phase Hamiltonian in the multipolar gauge, in the thermodynamic limit, is given by
\begin{eqnarray}
    H  & = & \hbar \sum_{\lambda, {\bf k}} \omega_k \Tilde{a}^\dag_{{\bf k}, \lambda} \Tilde{a}_{{\bf k}, \lambda} + \hbar \sum_{\alpha, {\bf k}} \left[ \omega_0 + 2 \frac{\Tilde{g}_k}{\Tilde{N}_k} \omega_k \sum_{\lambda, \beta} A_{{\bf k}, \lambda} B_{{\bf k}, \beta} e_{\lambda_{\beta}} - 4 \frac{\Tilde{g}^2_k}{\Tilde{N}_k} \omega_k \Tilde{f}_{{\bf k}, \beta, \gamma} \sum_{\beta, \gamma} B_{{\bf k}, \beta} B_{{\bf k}, \gamma} \right] \Tilde{b}^\dag_{{\bf k}, \alpha} \Tilde{b}_{{\bf k}, \alpha} \nonumber \\
    & + & i \hbar \sum_{\lambda, {\bf k}} \left[ 2 \Tilde{g}_k \omega_k \sum_\alpha B_{{\bf k}, \alpha} e_{\lambda_{\alpha}} - \omega_k A_{{\bf k}, \lambda} \right] \left( \Tilde{a}_{{\bf k}, \lambda} - \Tilde{a}^\dag_{-{\bf k}, \lambda} \right) \nonumber \\
    & + & \hbar \sum_{\alpha, {\bf k}} \left[ 2 \Tilde{g}_k \omega_k \sum_\lambda A_{{\bf k}, \lambda} e_{\lambda_{\alpha}} - \left( \omega_0 + 2 \frac{\Tilde{g}_k}{\Tilde{N}_k} \omega_k \sum_{\lambda, \beta} A_{{\bf k}, \lambda} B_{{\bf k}, \beta} e_{\lambda_{\beta}} \right) B_{{\bf k}, \alpha} \right. \nonumber \\
    & & \quad \quad \quad \left. + 4 \Tilde{g}^2_k \omega_k \left( \frac{B_{{\bf k}, \alpha}}{\Tilde{N}_k} \sum_{\beta, \gamma} B_{{\bf k}, \beta} B_{{\bf k}, \gamma} \Tilde{f}_{{\bf k}, \beta, \gamma} - \sum_{\beta} B_{{\bf k}, \beta} \Tilde{f}_{{\bf k}, \alpha, \beta} \right) \right] \left( \Tilde{b}_{{\bf k}, \alpha} + \Tilde{b}^\dag_{-{\bf k}, \alpha} \right) \nonumber \\
    & - & i \hbar \sum_{\alpha, \lambda, {\bf k}} \left[ \frac{\Tilde{g}_k}{\Tilde{N}_k} \omega_k \left( \Tilde{N}_k e_{\lambda_{\alpha}} - B_{{\bf k}, \alpha} \sum_\beta B_{{\bf k}, \beta} e_{\lambda_{\beta}} \right) \right] \left( \Tilde{a}_{{\bf k}, \lambda} - \Tilde{a}^\dag_{-{\bf k}, \lambda} \right) \left( \Tilde{b}_{-{\bf k}, \alpha} + \Tilde{b}^\dag_{{\bf k}, \alpha} \right) \nonumber \\
    & + & \hbar \sum_{\alpha, \beta, \lambda, {\bf k}} \left[ \frac{\Tilde{g}_k}{2 \Tilde{N}^2_k} \omega_k A_{{\bf k}, \lambda} B_{{\bf k}, \alpha} \left( 2 \Tilde{N}_k e_{\lambda_{\beta}} + B_{{\bf k}, \beta} \sum_\gamma B_{{\bf k}, \gamma} e_{\lambda_{\gamma}} \right) \right. \nonumber \\
    & & \quad \quad \quad \quad \left. + \frac{\Tilde{g}^2_k}{\Tilde{N}_k} \omega_k \left( \Tilde{N}_k \Tilde{f}_{{\bf k}, \alpha, \beta} - 4 B_{{\bf k}, \alpha} \sum_{\gamma} \Tilde{f}_{{\bf k}, \beta, \gamma} B_{{\bf k}, \gamma} \right) \right] \left( \Tilde{b}_{-{\bf k}, \alpha} + \Tilde{b}^\dag_{{\bf k}, \alpha} \right) \left( \Tilde{b}_{{\bf k}, \beta} + \Tilde{b}^\dag_{-{\bf k}, \beta} \right) \, , \label{app:H_M_SPT_full}
\end{eqnarray}
where $\Tilde{g}_k = g_k \sqrt{\Tilde{N}_k / N}$, $\Tilde{N}_k = N - \sum_\alpha B^2_{{\bf k}, \alpha}$ and $\Tilde{f}_{{\bf k}, \alpha, \beta} = \sum_{\lambda} e_{\lambda_{\alpha}} e_{\lambda_{\beta}} + f_{{\bf k}, \alpha, \beta}$.
The parameters $A_{{\bf k}, \lambda}$ and $B_{{\bf k}, \alpha}$, at equilibrium, are fixed by the stationary condition of the energy functional, which in turn is equivalent to impose the vanishing of the linear terms in the bosonic operators. Thus, for each mode $\bf k$, we have the resulting system of two coupled equations in the two parameters $A_{{\bf k}, \lambda}$ and $B_{{\bf k}, \alpha}$:
\begin{eqnarray}
    & 2 \Tilde{g}_k \omega_k \sum_\alpha B_{{\bf k}, \alpha} e_{\lambda_{\alpha}} - \omega_k A_{{\bf k}, \lambda} = 0 \nonumber \\
    & 2 \Tilde{g}_k \omega_k \left( \sum_\lambda A_{{\bf k}, \lambda} e_{\lambda_{\alpha}} - \frac{B_{{\bf k}, \alpha}}{\Tilde{N}_k} \sum_{\lambda, \beta} A_{{\bf k}, \lambda} B_{{\bf k}, \beta} e_{\lambda_{\beta}} \right) + 4 \Tilde{g}^2_k \omega_k \left( \frac{B_{{\bf k}, \alpha}}{\Tilde{N}_k} \sum_{\beta, \gamma} B_{{\bf k}, \beta} B_{{\bf k}, \gamma} \Tilde{f}_{{\bf k}, \beta, \gamma} - \sum_{\beta} B_{{\bf k}, \beta} \Tilde{f}_{{\bf k}, \alpha, \beta} \right) - \omega_0 B_{{\bf k}, \alpha} = 0. \nonumber \\
    &  \label{app:PT_sist_eq}
\end{eqnarray}

We now focus our attention on the transverse mode solutions, since those are the ones coupling with the radiation field.
This system of Eqs.~(\ref{app:PT_sist_eq}) admits, beside the trivial solution $A_{{\bf k}, \lambda} = B_{{\bf k}, \alpha} = 0$ corresponding to the normal phase where no condensation occurs, a nontrivial solution which is defined by the conditions for the parameters
\begin{eqnarray}
    \sum_\alpha B^2_{{\bf k}, \alpha} & = & \frac{N}{2} \left( 1 + \frac{1}{4 \eta^2 f^\perp_{\bf k} } \right) \, , \\
    A_{{\bf k}, \lambda} & = & 2 \Tilde{g}_k \sum_\alpha B_{{\bf k}, \alpha} e_{\lambda_{\alpha}} \, .
\end{eqnarray}

Inserting these values of the parameters in (\ref{app:H_M_SPT_full}), we obtain the Hamiltonian describing the condensed phase
\begin{eqnarray}
    H_{\rm cp}  & = & \hbar \sum_{\lambda, {\bf k}} \omega_k \Tilde{a}^\dag_{{\bf k}, \lambda} \Tilde{a}_{{\bf k}, \lambda} + \hbar \sum_{\alpha, {\bf k}} \omega_0 \frac{1 - 4 \eta^2 f^\perp_{\bf k}}{2} \Tilde{b}^\dag_{{\bf k}, \alpha} \Tilde{b}_{{\bf k}, \alpha} \nonumber \\
    & - & i \hbar \sum_{\alpha, \lambda, {\bf k}} \left[ \frac{\Tilde{g}_k}{\Tilde{N}_k} \omega_k \left( \Tilde{N}_k e_{\lambda_{\alpha}} - B_{{\bf k}, \alpha} \sum_\beta B_{{\bf k}, \beta} e_{\lambda_{\beta}} \right) \right] \left( \Tilde{a}_{{\bf k}, \lambda} - \Tilde{a}^\dag_{-{\bf k}, \lambda} \right) \left( \Tilde{b}_{-{\bf k}, \alpha} + \Tilde{b}^\dag_{{\bf k}, \alpha} \right) \nonumber \\
    & + & \hbar \sum_{\alpha, {\bf k}} \eta^2 \omega_0 \left(1 + f^\perp_{\bf k}\right) \frac{4 \eta^2 f^\perp_{\bf k} - 1}{8 \eta^2 f^\perp_{\bf k}} \left( \Tilde{b}_{-{\bf k}, \alpha} + \Tilde{b}^\dag_{{\bf k}, \alpha} \right) \left( \Tilde{b}_{{\bf k}, \alpha} + \Tilde{b}^\dag_{-{\bf k}, \alpha} \right) \nonumber \\
    & + & \hbar \sum_{\alpha, \beta, {\bf k}} \frac{\eta^2 \omega_0}{N} \left[ \frac{12 \eta^2 f^\perp_{\bf k} - 1}{4 \eta^2 f^\perp_{\bf k} - 1} - 4 \left(1 + f^\perp_{\bf k}\right) \right] B_{{\bf k}, \alpha} B_{{\bf k}, \beta} \left( \Tilde{b}_{-{\bf k}, \alpha} + \Tilde{b}^\dag_{{\bf k}, \alpha} \right) \left( \Tilde{b}_{{\bf k}, \beta} + \Tilde{b}^\dag_{-{\bf k}, \beta} \right) \, . \label{app:H_M_condensed}
\end{eqnarray}

\begin{figure}[t]
    \centering
    \includegraphics[width=0.6\linewidth]{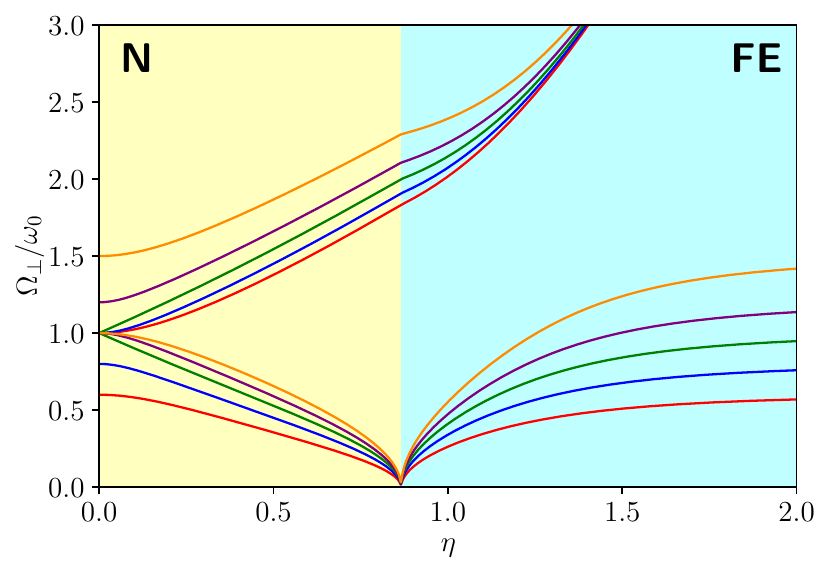}
    \caption{Upper and lower polaritons as functions of $\eta$ for different modes: $\omega_k / \omega_0 = 0.6$ (red), $0.8$ (blue), $1$ (green), $1.2$ (purple), $1.5$ (yellow). The polaritons in the condensed ferroelectric phase are calculated by Eq.~(\ref{app:disp_rel_post}).}
    \label{fig:disp_rel_up_low}
\end{figure}

Calculating the dispersion relations induced by this Hamiltonian (\ref{app:H_M_condensed}) as outlined in Sec. \ref{sec:disp_rel}, we obtain the following dispersion relation for the transverse mode solutions
\begin{equation}
    \frac{\omega^2_{\bf k}}{\Omega^2_\perp} = 1 + \frac{ \omega^2_0 / f^\perp_{\bf k} }{ \omega^2_0 \left( 1 - 16 \eta^4 {f_{\bf k}^\perp}^2 \right) + \Omega^2_\perp } \, . \label{app:disp_rel_post}
\end{equation}
Therefore, combining Eqs.~(\ref{app:disp_rel_pre}) and (\ref{app:disp_rel_post}) for the dispersion relations before and after the QPT, we obtain (see Fig.~\ref{fig:disp_rel_up_low})
\begin{equation}
    \begin{cases}
        \frac{\omega^2_{\bf k}}{\Omega^2_\perp} = 1 + \frac{4 \eta^2 \omega^2_0}{\omega^2_0 \left( 1 + 4 \eta^2 f^\perp_{\bf k} \right) - \Omega^2_\perp} & \eta < \eta_c \, , \\
        \frac{\omega^2_{\bf k}}{\Omega^2_\perp} = 1 + \frac{ \omega^2_0 / f^\perp_{\bf k} }{ \omega^2_0 \left( 1 - 16 \eta^4 {f_{\bf k}^\perp}^2 \right) + \Omega^2_\perp } & \eta > \eta_c \, .
    \end{cases}
    \label{app:disp_rel_bulk_transv}
\end{equation}

We can now investigate the ground state condensation occurring after the QPT. This can be easily understood in terms of mean mode occupation of the physical fields: in particular the electric displacement field $\bf D (r)$ can be easily demonstrated to have a mean mode occupation, $\langle D_{\bf k} \rangle$, which equals to the mean transverse polarization field $\langle P^\perp_{\bf k} \rangle$. Thus, given the definition of the displacement field ${\bf D(r)} = \epsilon_m {\bf E^\perp (r)} + {\bf P^\perp (r)}$, the property $\langle D_{\bf k} \rangle = \langle P^\perp_{\bf k} \rangle$ implies that the mean mode occupation of the electric field, $\langle E^\perp_{\bf k} \rangle$, is zero. Moreover, we remark that the same mean field mode occupations are predicted even in the two-step Bogoliubov diagonalization, further confirming the ferroelectricity of the QPT. In fact, the mean occupation of the matter field is dictated only by the strength of dipolar interactions and not on the interaction with the radiation field.

As a final note, it is instructive to point out that if the factor $f_{{\bf k}, \alpha, \beta}$ is regarded as a free parameter, we can recover from Hamiltonians (\ref{app:H_M_bulk_full}) and (\ref{app:H_M_SPT_full}) relevant known models and their behaviors near phase transitions. For instance, the Dicke model can be recovered by restricting the study only to the transverse modes and considering $f^\perp_{{\bf k}, \alpha, \beta} = - \delta_{\alpha \beta}$ (implying $\Tilde{f}^\perp_{{\bf k}, \alpha, \beta} = 0$). As a consequence, relation (\ref{app:disp_rel_post}) reduces to the already known Dicke dispersion relation above the SPT \cite{Brandes03}. Another notable case is $f^\perp_{{\bf k}, \alpha, \beta} = 0$, corresponding to a pure Hopfield model, neglecting the dipole-dipole interactions. For such a model, the system of equations (\ref{app:PT_sist_eq}) admits only the trivial solution, which is consistent with the well-known impossibility
for a pure Hopfield Hamiltonian to undergo a phase transition.

\section{Connection to experimental data} \label{sec:experimental_data}

\begin{figure}[b]
    \centering
    \includegraphics[width=\textwidth]{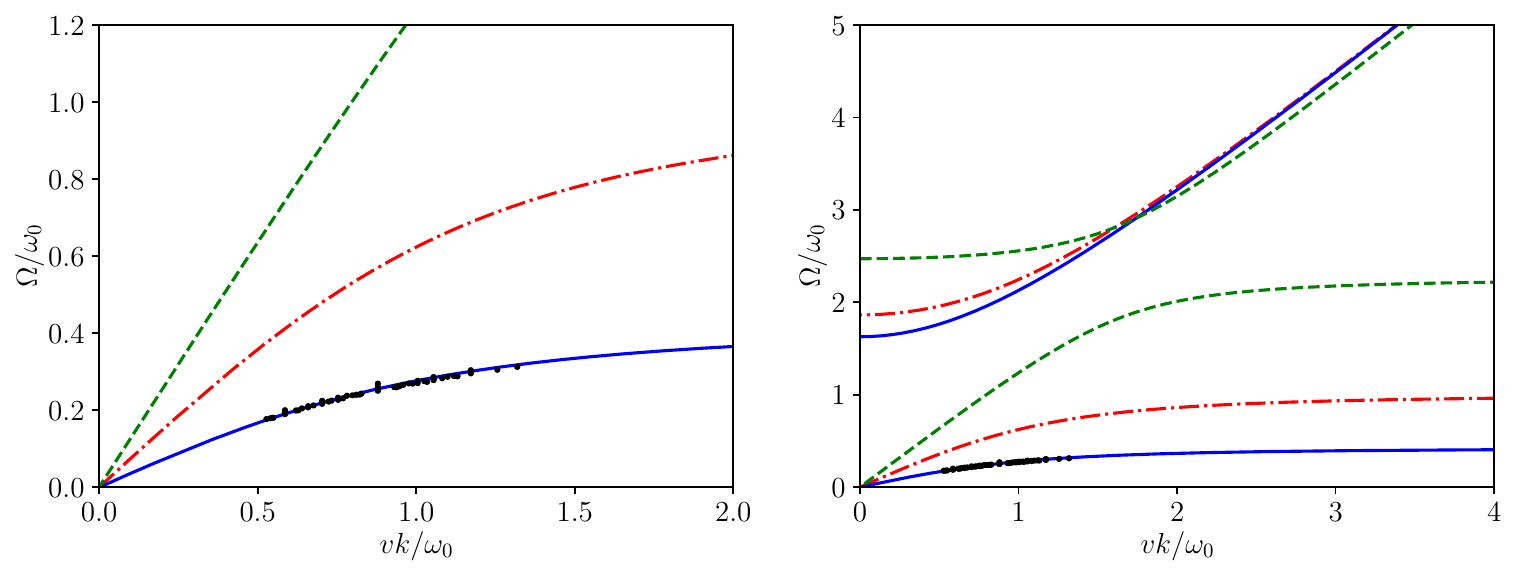}
    \caption{The left figure is focused on the region near the experimental data, while the right figure presents a broader view including the upper polariton. As shown, the theoretical dispersion relation derived from the Hopfield-like model including the dipole-dipole interactions (solid blue) perfectly fits the experimental data (black dots). In contrast, the curves derived from the Dicke-like model (dashed green) and from the Hopfield model neglecting the dipolar interaction (red dash-dotted) significantly differ from the experimental data.
    The relevant parameters to the theoretical curves are $\eta' = 1.83$, $\omega_0 = 1.83 \; eV$ and $\epsilon_m = 1.96$, while the full set of parameters for the experimental curves are found in the respective reference \cite{Nature_exp_nanoparticles}.}
    \label{fig:extended_data}
\end{figure}
Here we further analyze and discuss the differences in the models in light of the experimental data presented in Ref.~\cite{Nature_exp_nanoparticles}, which refer to gold nanoparticle crystals with normalized light-matter coupling strength in the deep-strong coupling regime. Such a system falls under the model described in the main text, given the highly localized particles whose polarizations fields can be safely considered non-overlapping. In Fig. \ref{fig:extended_data}, we present the comparison between the dispersion relations derived by a Dicke-like model, a Hopfield-like model including the dipole-dipole interactions and a Hopfield-like model neglecting these interactions, for renormalized light-matter coupling $\eta' = 1.83$, corresponding to $\eta = 0.78$. The dispersion curves differ substantially for such high couplings: while the Hopfield model neglecting the dipolar interactions does not predict any QPT, the Hopfield-like model including this term is in the neighborhood of the ferroelectric QPT, whereas in the Dicke-like model the SPT has already taken place. Experimental measurements regarding the lower polariton are reported in the figure (black dots), which perfectly fits the theoretical predictions of the Hopfield-like model including the dipolar interactions, thus demonstrating its validity. On the other hand, these measurements are incompatible with the predictions of the other models. No data is available for the upper polariton given its high energy, which collocates it above the onset of the gold interband transitions.

\end{document}